\documentclass[conference]{IEEEtran}
\IEEEoverridecommandlockouts
\makeatletter
\def\ps@headings{%
\def\@oddhead{\mbox{}\scriptsize\rightmark \hfil \thepage}%
\def\@evenhead{\scriptsize\thepage \hfil \leftmark\mbox{}}%
\def\@oddfoot{}%
\def\@evenfoot{}}
\makeatother
\usepackage{cite}
\usepackage{boxedminipage}
\usepackage{amsthm}
\usepackage{amssymb,amsmath,bm}
\usepackage{multirow}
\usepackage{multicol}
\usepackage{macros}
\usepackage{times}
\usepackage{graphicx}
\usepackage{epsfig}
\usepackage{color}
\usepackage{array}
\usepackage{tabularx}
\usepackage{wrapfig}
\usepackage{subfig}
\usepackage{algorithm}
\usepackage{algorithmic}

\ifCLASSINFOpdf
\else
\fi

\usepackage{url}

\begin{document}
\title{Secure Multidimensional Queries in Tiered Sensor Networks}

\author{\IEEEauthorblockN{Chia-Mu Yu\IEEEauthorrefmark{2}\IEEEauthorrefmark{4}, Chun-Shien Lu\IEEEauthorrefmark{2}, and Sy-Yen Kuo\IEEEauthorrefmark{4}}

\IEEEauthorblockA{\IEEEauthorrefmark{2}Institute of Information Science, Academia Sinica, Taipei, Taiwan}
\IEEEauthorblockA{\IEEEauthorrefmark{4}Department of Electrical Engineering, National Taiwan University, Taipei, Taiwan}
}

\maketitle

\begin{abstract}
In this paper, aiming at securing range query, top-k query, and skyline query in tiered sensor networks, we propose the Secure Range Query (SRQ), Secure Top-$k$ Query (STQ), and Secure Skyline Query (SSQ) schemes, respectively. In particular, SRQ, by using our proposed \emph{prime aggregation} technique, has the lowest communication overhead among prior works, while STQ and SSQ, to our knowledge, are the first proposals in tiered sensor networks for securing top-$k$ and skyline queries, respectively. Moreover, the relatively unexplored issue of the security impact of sensor node compromises on multidimensional queries is studied; two attacks incurred from the sensor node compromises, \emph{collusion attack} and \emph{false-incrimination attack}, are investigated in this paper. After developing a novel technique called \emph{subtree sampling}, we also explore methods of efficiently mitigating the threat of sensor node compromises. Performance analyses regarding the probability for detecting incomplete query-results and communication cost of the proposed schemes are also studied.
\end{abstract}

\section{Introduction}\label{sec: Introduction}

\textbf{Tiered Sensor Networks.} Sensor networks are expected to be deployed on some harsh or hostile regions for data collection or environment monitoring. Since there is the possibility of no stable connection between the authority and the network, in-network storage is necessary for caching or storing the data sensed by sensor nodes. A straightforward method is to attach external storage to each node, but this is economically infeasible. Therefore, various data storage models for sensor networks have been studied in the literature. In \cite{rksegyy03,dglls05}, a notion of tiered sensor networks was discussed by introducing an intermediate tier between the authority and the sensor nodes. The purpose of this tier is to cache the sensed data so that the authority can efficiently retrieve the cache data, avoiding unnecessary communication with sensor nodes.

The network model considered in this paper is the same as the ones in \cite{rksegyy03,dglls05}. More specifically, some storage-abundant nodes, called \emph{storage nodes}, which are equipped with several gigabytes of NAND flash storage \cite{slm06}, are deployed as the intermediate tier for data archival and query response. In practice, some currently available sensor nodes such as RISE \cite{RISE} and StarGate \cite{startgate} can work as the storage nodes. The performance of sensor networks wherein external flash memory is attached to the sensor nodes was also studied in \cite{mdgs06}. In addition, some theoretical issues concerning the tiered sensor networks, such as the optimal storage node placement, were also studied in \cite{slm06,stlm07}. In fact, such a two-tiered network architecture has been demonstrated to be useful in increasing network capacity and scalability, reducing network management complexity, and prolonging network lifetime.

\textbf{Multidimensional Queries.} Although a large amount of sensed data can be stored in storage nodes, the authority might be interested in only some portions of them. To this end, the authority issues proper queries to retrieve the desired portion of sensed data. Note that, when the sensed data have multiple attributes, the query could be multidimensional. We have observed that range query, top-$k$ query, and skyline query are the most commonly used queries. Range query \cite{lkgh03,mfhh05}, which could be useful for correlating events occurring within the network, is used to retrieve sensed data whose attributes are individually within a specified range. After mapping the sensed data to a ranking value, top-$k$ query \cite{wxtl07}, which can be used to extract or observe the extreme phenomenon, is used to retrieve the sensed data whose ranking values are among the first $k$ priority. Skyline query \cite{czg07,lcy08}, due to its promising application in multi-criteria decision making, is also useful and important in environment monitoring, industry control, \emph{etc}.

Nonetheless, in the tiered network model, the storage nodes become the targets that are easily compromised because of their significant roles in responding to queries. For example, the adversary can eavesdrop on the communications among nodes or compromise the storage nodes to obtain the sensed data, resulting in the breach of \emph{data confidentiality}. After the compromise of storage nodes, the adversary can also return falsified query-results to the authority, leading to the breach of \emph{query-result authenticity}. Even more, the compromised storage nodes can cause \emph{query-result incompleteness}, creating an incomplete query-result for the authority by dropping some portions of the query-result.

\textbf{Related Work.} Secure range queries in tiered sensor networks have been studied only in \cite{sl08,szz09,zsz09}. Data confidentiality and query-result authenticity can be preserved very well in \cite{sl08,szz09,zsz09} owing to the use of the bucket scheme \cite{hilm02,hmt04}. Unfortunately, encoding approach \cite{sl08} is only suitable for the one-dimensional query scenario in the sensor networks for environment monitoring purposes. On the other hand, crosscheck approaches \cite{szz09,zsz09} can be applied on sensor networks for event-driven purposes at the expense of the reduced probability for detecting query-result incompleteness. The security issues incurred from the compromise of storage nodes have been addressed in \cite{sl08,szz09,zsz09}. The impact of \emph{collusion attacks} defined as the collusion among compromised sensor nodes and compromised storage nodes, however, was only discussed in \cite{zsz09}, wherein only a naive method was proposed as a countermeasure. When the compromised sensor nodes are taken into account, a Denial-of-Service attack, called \emph{false-incrimination attack}, not addressed in the literature, can be extremely harmful. In such an attack, the compromised sensor nodes subvert the functionality of the secure query schemes by simply claiming that their sensed data have been dropped by the storage nodes. After that, the innocent storage nodes will be considered compromised and will be revoked by the authority. It should be noted that all the previous solutions suffer from false-incrimination attacks.

\textbf{Contribution.} Our major contributions are:
\begin{itemize}
\item The Secure Range Query (SRQ) scheme is proposed to secure the range query in tiered networks (Sec. \ref{sec: Securing Range Queries (SRQ)}). By taking advantage of our proposed \emph{prime aggregation} technique for securely transmitting the amount of data in specified buckets, SRQ has the lowest communication cost among prior works in all scenarios (environment monitoring and event detection purposes), while preserving the probability for detecting incomplete query-results close to $1$. It should be noted that although incorporating bucket scheme \cite{hilm02,hmt04} (described in Sec. \ref{sec: Confidentiality-preserving reporting}) in the protocol design \cite{sl08,szz09,zsz09} is not new, the novelty of our method lies on the use of \emph{prime aggregation} in reducing the overhead and guaranteeing query-result completeness.

\item For the first time in the literature, the issues of securing top-$k$ and skyline queries in tiered networks are studied (Secs. \ref{sec: Securing Top-$k$ Queries (STQ)} and \ref{sec: Securing Skyline Queries (SSQ)}). Our solutions to these two issues are Secure Top-$k$ Query (STQ) and Secure Skyline Query (SSQ), respectively. The former is built upon the proposed SRQ scheme to detect query-result completeness, while the efficiency of SSQ is based on our proposed \emph{grouping} technique.

\item The security impact of sensor compromises is studied (Sec. \ref{sec: Impact of Sensor Node Compromise}); \emph{collusion attack} is formally addressed, and a new Denial-of-Service attack, \emph{false-incrimination attack}, which can thwart the security purpose in prior works, is first identified in our paper. The resiliency of SRQ, STQ, and SSQ against these two attacks is investigated. With a novel technique called \emph{subtree sampling}, some minor modifications are introduced for SRQ and STQ as countermeasures to these two attacks. Moreover, the compromised nodes can even be efficiently identified and be further attested \cite{slpdk06,smkk05,spdk04}.
\end{itemize}
\section{System Model}\label{sec: System Model}
In general, the models used in this paper are very similar to those in \cite{rksegyy03,dglls05,sl08,szz09,zsz09}.

\begin{wrapfigure}{r}{0.25\textwidth}
  \begin{center}
    \includegraphics[width=0.32\textwidth]{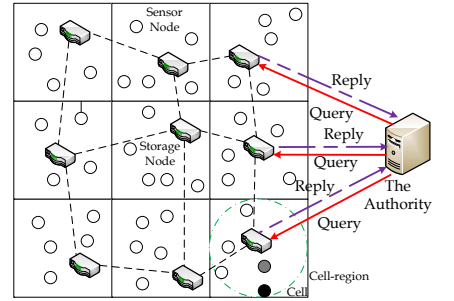}
  \end{center}
  \caption{\footnotesize A tiered sensor network.}
  \label{fig: tiered_network}
\end{wrapfigure}

\textbf{Network Model.} As shown in Fig. \ref{fig: tiered_network}, the sensor network considered in this paper is composed of a large number of resource-constrained sensor nodes and a few so-called \emph{storage nodes}. Storage nodes are assumed to be storage-abundant and may be compromised. In addition, in certain cases, storage nodes could also have abundant resources in energy, computation, and communication. The storage nodes can communicate with the authority via direct or multi-hop communications. The network is connected such that, for two arbitrary nodes, at least one path connecting them can be found.

A \emph{cell} is composed of a storage node and a number of sensor nodes. In a cell, sensor nodes could be far away from the associated storage node so that they can communicate with each other only through multi-hop communication. For example, in Fig. \ref{fig: tiered_network}, without the relay of the gray node, the black node cannot reach the storage node.  The nodes in the network have synchronized clocks \cite{snwlz06} and the time is divided into epochs. As in \cite{sl08,slm06,szz09,zsz09}, each node is assumed to be aware of the geographic position it locates \cite{lnd05,zlfw06} so that the association between the sensor node and storage node can be established. As a matter of fact, information about the time and geographic position is indispensable for most sensor network applications.

For each cell, aggregation is assumed to be performed over an aggregation tree rooted at the storage node. Since the optimization of the aggregation tree structure is out of the scope of this paper, we adopt the method described in TAG \cite{mfhh02} to construct an aggregation tree. We follow the conventional assumption that the topology of the aggregation tree is known by the authority \cite{cp08,cps06}. Similar to sensor nodes, storage nodes also perform the sensing task. Each sensor node senses the data and temporarily stores the sensed data in its local memory within an epoch. At the end of each epoch, the sensor nodes in a cell report the sensed data stored in local memory to the associated storage node. Throughout this paper, we focus on a cell $\mathcal{C}$, composed of $N-1$ sensor nodes, $\{s_i\}_{i=1}^{N-1}$, and a storage node $\mathcal{M}$.

\textbf{Security Model.}
We consider the adversary who can compromise an arbitrary number of storage nodes. After node compromises, all the information stored in the compromised storage nodes will be exposed to the adversary. The goal of the adversary is to breach at least one of the following: data confidentiality, query-result authenticity, and query-result completeness. We temporarily do not consider the compromise of sensor nodes in describing SRQ, STQ, and SSQ in Sec. \ref{sec: Securing Multidimensional Queries}. The impact of sensor node compromise on the security breach, however, will be explored in Sec. \ref{sec: Impact of Sensor Node Compromise}. Many security issues in sensor networks, such as key management \cite{cps03,eg02,ylk09}, broadcast authentication \cite{pswct01,ln04}, and secure localization \cite{lnd05,zlfw06}, have been studied in the literature. This paper focuses on securing multidimensional queries that are relatively unexplored in the literature, while the protocol design of the aforementioned issues are beyond the scope of this paper.

\textbf{Query Model.}
The sensed data can be represented as a $d$-dimensional tuple, $(A_1,A_2,\dots, A_d)$, where $A_g$, $\forall g\in[1,d]$, denotes the $g$-th attribute. The authority may issue a proper $d$-dimensional query to retrieve the desired portion of data stored in storage nodes. Three types of queries, including range query, top-$k$ query, and skyline query, are considered in this paper. For range query, its form, issued by the authority, is expressed as $\langle \mathcal{C}, t, l_1, h_1, \dots, l_d, h_d\rangle$, which means that the sensed data to be reported to the authority should be generated by the nodes in cell $\mathcal{C}$ at epoch $t$, and their $g$-th attributes, $A_g$'s, should be within the range of $[l_g,h_g]$, $g\in[1,d]$. Top-$k$ query is usually associated with a scalar (linear) \emph{ranking function}. With ranking function, $R$, the sensed data, even if it is multidimensional, can be individually mapped to a one-dimensional \emph{ranking value}. The top-$k$ query issued by the authority is in the form of $\langle \mathcal{C}, t, R, k\rangle$. As the first attempt to achieve secure top-$k$ query, the goal of top-$k$ query in this paper is simply assumed to obtain the sensed data generated by the nodes in cell $\mathcal{C}$ at epoch $t$ with the first $k$ smallest ranking values. For skyline query, the desired \emph{skyline data} are defined as those not \emph{dominated} by any other data. Assuming that smaller values are preferable to large ones for all attributes, for a set of $d$-dimensional data, a datum $c_i$ \emph{dominates} another datum $c_j$ if both the conditions, $A_g(c_i)\leq A_g(c_j)$, $\forall g\in [1,d]$, and $A_g(c_i)< A_g(c_j), \exists g\in [1,d]$, where $A_g(c_i)$ denotes the $g$-th attribute value of the datum $c_i$, hold. Hence, the form of the skyline query issued by the authority is given as $\langle \mathcal{C}, t\rangle$, which is used to retrieve the skyline data generated in cell $\mathcal{C}$ at epoch $t$.

\section{Securing Multidimensional Queries}\label{sec: Securing Multidimensional Queries}
In this section, aiming at securing range query, top-$k$ query, and skyline query, we propose the SRQ (Sec. \ref{sec: Securing Range Queries (SRQ)}), STQ (Sec. \ref{sec: Securing Top-$k$ Queries (STQ)}), and SSQ (Sec. \ref{sec: Securing Skyline Queries (SSQ)}) schemes, respectively. Note that though SRQ, STQ, and SSQ use the bucket scheme \cite{hilm02,hmt04}, the novelty of them is due to their design in efficiently detecting the incomplete query-result (described later).

\subsection{Securing Range Queries (SRQ)}\label{sec: Securing Range Queries (SRQ)}
Our proposed SRQ scheme consists of a confidentiality-preserving reporting phase (Sec. \ref{sec: Confidentiality-preserving reporting}) that can simultaneously prevent the adversary from accessing data stored in the storage nodes, authenticate the query results, and ensure efficient multidimensional query processing, and a query-result completeness verification phase (Sec. \ref{sec: Query-Result Completeness Verification}) for guaranteeing the completeness of query-results.

\subsubsection{Confidentiality-preserving reporting}\label{sec: Confidentiality-preserving reporting}
Data encryption is a straightforward and common method of ensuring data confidentiality against a compromised storage node. Moreover, we hope that even when the adversary compromises the storage node, the previously stored information should not be exposed to the adversary. To this end, the keys used in encryption should be selected from a one-way hash chain. In particular, assume that a key $K_{i,0}$ is initially stored in sensor node $s_i$. At the beginning of epoch $t$, the key $K_{i,t}$, which is used only within epoch $t$, is calculated as $hash(K_{i,t-1})$, where $hash(\cdot)$ is a hash function, and $K_{i,t-1}$ is dropped. Suppose that sensor node $s_i$ has sensed data $D$ at epoch $t$. One method for storing $D$ in the storage node $\mathcal{M}$ while preserving the privacy is to send $\{D\}_{K_{i,t}}$, which denotes the encryption of $D$ with the key $K_{i,t}$. With this method, when an OCB-like authenticated encryption primitive \cite{rbb03} is exploited, the authenticity of $D$ can be guaranteed. At the same time, $D$ will not be known by the adversary during message forwarding and even after the compromise of the storage node at epoch $t$ because the adversary cannot recover the keys used in the time before epoch $t$. Nevertheless, no query can be answered by $\mathcal{M}$ if only encrypted data is stored in $\mathcal{M}$. Hence, the \emph{bucket scheme} proposed in \cite{hilm02,hmt04}, which uses the encryption keys generated via a one-way hash chain, is used in the SRQ scheme.

In the bucket scheme, the domain of each attribute $A_g$, $\forall g \in [1,d]$, is assumed to be known in advance, and is divided into $w_g\geq 1$ consecutive non-overlapping intervals sequentially indexed from $1$ to $w_g$, under a publicly known partitioning rule. For ease of representation, in the following, we assume that $w_g=w$, $\forall g \in[1,d]$. A $d$-dimensional bucket is defined as a tuple, $(v_1, v_2,\dots, v_d)$ (hereafter called \emph{bucket ID}), where $v_g\in [1,w]$, $g\in [1,d]$. The sensor node $s_i$, when it has sensed data at epoch $t$,  sends to $\mathcal{M}$ the corresponding bucket IDs, which are constructed by mapping each attribute of the sensed data to the proper interval index, and the sensed data encrypted by the key $K_{i,t}$. For example, when $s_i$ has sensed data $(1,3)$, $(2,4)$, and $(2,11)$ at epoch $t$, the message transmitted to the storage node at the end of epoch $t$ is $\langle i,t,(1,1),\{(1,3),(2,4)\}_{K_{i,t}},(1,2),\{(2,11)\}_{K_{i,t}}\rangle$, assuming that $A_1, A_2\in [1,20]$, $w=2$, and each interval length, set at $10$, is the same.

Let $\mathcal{V}$ be the set of all possible bucket IDs. Assume that there are on average $Y$ and $Y/N$ data generated in a cell and in a node, respectively, at epoch $t$. Assume that, $D_{i,t,V}$ is a set containing all the data within the bucket $V\in\mathcal{V}$ sensed by $s_i$ at epoch $t$. The messages sent from $s_i$ to $\mathcal{M}$ at the end of epoch $t$ can be abstracted as $\langle i,t,J_\sigma,\{D_{i,t,J_\sigma}\}_{K_{i,t}}\rangle$, where $J_\sigma\in \mathcal{V}, J_\sigma\neq J_{\sigma'}, 1\leq \sigma,\sigma'\leq Y/N$ if there are some data sensed by $s_i$ within epoch $t$. Note that $s_i$ sends nothing to $\mathcal{M}$ if $D_{i,t,J_\sigma}$'s, $\forall J_{\sigma}\in \mathcal{V}$, are empty. After that, $\mathcal{M}$ can answer the range query according to the information revealed by the bucket IDs. Assume that $l_g$ and $h_g$ are located within the $\alpha_g$-th and $\beta_g$-th intervals, respectively, where $\alpha_g\leq \beta_g$, $\alpha_g$, $\beta_g\in [1,w]$, and $g\in [1,d]$. The encrypted data falling into the buckets in the set $\mathcal{A}=\{(\rho_1,\dots,\rho_d)|\alpha_g \leq \rho_g\leq \beta_g, g\in[1,d]\}$ are reported to the authority. In other words, once receiving the range query, $\mathcal{M}$ first translates the information $l_1, h_1, \dots, l_d, h_d$ into the proper bucket IDs and then replies all the encrypted data falling into the buckets\footnote{There is a tradeoff between the communication cost and confidentiality in terms of bucket sizes because larger bucket size implies higher data confidentiality and higher communication cost due to more superfluous data being returned to the authority. The design of optimal bucketing strategies is beyond the scope of this paper, and we refer to \cite{hilm02,hmt04} for more details.} in $\mathcal{A}$.

Nevertheless, in tiered sensor networks, even when the original bucket scheme is used, $\mathcal{M}$ could still maliciously drop some encrypted data and only report part of the results to the authority, resulting in an incomplete query-result. In the following, we will describe an extended bucket scheme, which incorporates the prime aggregation strategy into the original bucket scheme, to detect the incomplete reply in a communication-efficient manner.

\subsubsection{Query-Result Completeness Verification}\label{sec: Query-Result Completeness Verification}
With \emph{prime aggregation} technique, SRQ detects an incomplete reply by taking advantage of aggregation for counting the amount of sensed data falling into specified buckets. Together with a hash for verification purpose, the count forms a so-called \emph{proof} in detecting an incomplete reply. The storage node $\mathcal{M}$ is required to provide the proof to the authority at the epoch specified in the query so that the authority can use the proof to verify the completeness of received query-results. Since in our design all the sub-proofs generated by the nodes can be aggregated to yield the final proof, the communication cost can be significantly reduced. The details are described as follows.

Assume that an aggregation tree \cite{mfhh02} has been constructed after sensor deployment. Recall that the domain of attribute $A_g$ is divided into $w$ intervals. Before the sensor deployment, a set $\{p_{i}^{V},p_{i}^{\emptyset}|\forall i\in[1,N-1],V\in\mathcal{V}\}$ of $(wd+1)(N-1)$ prime numbers is selected by the authority such that $p_{i}^V\neq p_{i'}^{V'}$ and $p_{i}^{\emptyset}\neq p_{i'}^{\emptyset}$ if $i\neq i'$ or $V\neq V'$. Then, the set $\{p_{i}^{V},p_{i}^{\emptyset}| V\in\mathcal{V}\}$ of $wd+1$ prime numbers, called the set of \emph{bucket primes} of $s_i$, is stored in each sensor node $s_i$. In addition, a set $\{k_{i,0}^{V},k_{i,0}^{\emptyset}|V\in\mathcal{V}\}$ of $wd+1$ keys is selected by the authority and is stored in each sensor node $s_i$ initially. For fixed $i$ and $t$, the set of $\{k_{i,t}^{V},k_{i,0}^{\emptyset}|V\in\mathcal{V}\}$ is called the set of \emph{bucket keys} of $s_i$ at epoch $t$. Bucket primes could be publicly-known, while bucket keys should be kept secret. Each sensor node $s_i$, at the beginning of epoch $t$, calculates $k_{i,t}^{V}=hash(k_{i,t-1}^{V})$ and then drops $k_{i,t-1}^{V}$, $\forall V\in\mathcal{V}$. In addition, $s_i$ also calculates $k_{i,t}^{\emptyset}=hash(k_{i,t-1}^{\emptyset})$ and then drops $k_{i,t-1}^{\emptyset}$.

Recall that each node $s_i$ on average has $Y/N$ sensed data at epoch $t$, and assume that the set of $Y/N$ bucket IDs associated with these $Y/N$ sensed data is $B_{i,t}=\{\hat{v}^{i,t,\sigma}|\sigma=1,\dots, Y/N\}$, which could be a multiset. Then, according to its sensed data, $s_{i}$ calculates $H_{i,t}=hash_{K_{i,t}}(\sum_{\sigma=1}^{Y/N}k_{i,t}^{\hat{v}^{i,t,\sigma}})$, where $hash_{K}(\cdot)$ denotes the keyed hash function with key $K$, if it has sensed data, and $H_{i,t}=hash_{K_{i,t}}(k_{i,t}^{\emptyset})$ otherwise. Moreover, $s_i$ computes $P_{i,t}=\prod_{\sigma=1}^{Y/N}p_{i}^{\hat{v}^{i,t,\sigma}}$ if it has sensed data, and $P_{i,t}=p_i^{\emptyset}$ otherwise. Moreover, once receiving $\langle j_\rho, t, \mathcal{E}_{j_\rho,t}, \mathcal{B}_{j_\rho,t}, \mathcal{H}_{j_\rho,t}, \mathcal{P}_{j_\rho,t}\rangle, j_{\rho}\in[1,N-1]$, $\forall \rho\in [1,\chi]$ from its $\chi$ children, $s_{j_1},\dots,s_{j_\chi}$, $s_{i}$ calculates $\mathcal{E}_{i,t}=(\bigcup_{\rho=1}^{\chi}\mathcal{E}_{j_{\rho},t})\bigcup E_{i,t}$, where $E_{i,t}$ denotes the set of encrypted data sensed by $s_i$ at epoch $t$, and $\mathcal{B}_{i,t}=(\bigcup_{\rho=1}^{\chi}\mathcal{B}_{j_{\rho},t})\bigcup B_{i,t}$, where $B_{i,t}$ denotes the set of bucket IDs of $E_{i,t}$. In addition, $s_{i}$ also calculates $\mathcal{H}_{i,t}=hash_{K_{i,t}}(\sum_{\rho=1}^\chi \mathcal{H}_{j_\rho,t}+H_{i,t})$ and $\mathcal{P}_{i,t}=\prod_{\rho=1}^\chi \mathcal{P}_{j_\rho,t}\cdot P_{i,t}$, $\rho\in [1,\chi]$. Finally, $s_{i}$ reports $\langle i, t, \mathcal{E}_{i,t}, \mathcal{B}_{i,t}, \mathcal{H}_{i,t}, \mathcal{P}_{i,t}\rangle$ to its parent node on the aggregation tree. Note that, if $s_i$ is a leaf node on the aggregation tree, then we assume that it receives $\langle \emptyset,\emptyset,\emptyset,\emptyset,0,1\rangle$.

Assume that the set $\{p_{\mathcal{M}}^{V},p_{\mathcal{M}}^{\emptyset}|V\in\mathcal{V}\}$ of $wd+1$ prime numbers stored in $\mathcal{M}$ are all different from those stored in sensor nodes, and the set $\{k_{\mathcal{M},0}^{V},k_{\mathcal{M},0}^{\emptyset}|V\in\mathcal{V}\}$ of $wd+1$ bucket keys are selected by the authority and stored in $\mathcal{M}$. $\mathcal{M}$ computes $k_{\mathcal{M},t}^{V}=hash(k_{\mathcal{M},t-1}^{V})$ and drops $k_{\mathcal{M},t-1}^{V}$ at epoch $t$. In addition, $\mathcal{M}$ also computes $k_{\mathcal{M},t}^{\emptyset}=hash(k_{\mathcal{M},t-1}^{\emptyset})$ and drops $k_{\mathcal{M},t-1}^{\emptyset}$ at epoch $t$. For the storage node $\mathcal{M}$, it can also calculate $E_{\mathcal{M},t}$, $B_{\mathcal{M},t}$, $H_{\mathcal{M},t}$, and $P_{\mathcal{M},t}$ according to the its own sensed data at epoch $t$. In fact, the procedures $\mathcal{M}$ needs to perform after messages are received from the child nodes are the same as the ones performed by the sensor nodes. Acting as the root of the aggregation tree, however, $\mathcal{M}$ keeps the aggregated results, which are denoted as $\mathcal{E}_{\mathcal{M},t},\mathcal{B}_{\mathcal{M},t},\mathcal{H}_{\mathcal{M},t}$ and $\mathcal{P}_{\mathcal{M},t}$, respectively, in its local storage and waits for the query issued by the authority. Note that $P_{\mathcal{M},t}$ can be thought of as a compact summary of the sensed data of the whole network and can be very useful for the authority in checking the completeness of the query-result, while $H_{\mathcal{M},t}$ can be used by the authority to verify the authenticity of $P_{\mathcal{M},t}$.

Assume that a range query $\langle \mathcal{C}, t, l_1, h_1, l_2, h_2, \dots, l_d, h_d\rangle$ is issued by the authority. The encrypted data falling into the buckets in the set $\mathcal{A}$, along with the proof composed of $\mathcal{H}_{\mathcal{M},t}$ and $\mathcal{P}_{\mathcal{M},t}$, are sent to the authority. Once $\mathcal{P}_{\mathcal{M},t}$ is received, the authority immediately performs the prime factor decomposition of $\mathcal{P}_{\mathcal{M},t}$. Due to the construction of $\{p_{i}^{V},p_{\mathcal{M}}^{V},p_{i}^{\emptyset},p_{\mathcal{M}}^{\emptyset}|i\in[1,N-1],V\in\mathcal{V}\}$, which guarantees that the bucket primes are all distinct, after the prime factor decomposition of $\mathcal{P}_{\mathcal{M},t}$, the authority can be aware of which node contributes which data within specified buckets. As a result, the authority can know which keys should be used to verify the authenticity and integrity of $\mathcal{H}_{\mathcal{M},t}$. More specifically, assume that $\mathcal{P}_{\mathcal{M},t}=(\mathfrak{p}_1)^{a_1}\cdots (\mathfrak{p}_\gamma)^{a_\gamma}, a_1,\dots,a_\gamma> 0$, $\gamma\geq 0$, and that $\mathfrak{p}_1,\dots,\mathfrak{p}_{\gamma}$ are distinct prime numbers. From the construction of $P_{\mathcal{M},t}$, we know that $(\mathfrak{p}_{\hat{k}})^{a_{\hat{k}}}$, for $\hat{k}\in[1,\gamma]$, is equal to $(p_{k'}^{k''})^{a_{\hat{k}}}$, for $k'\in[1,N-1]$ and $k''\in \mathcal{V}$. From the procedure performed by each node, it can also be known that the appearance of $(\mathfrak{p}_{\hat{k}})^{a_{\hat{k}}}=(p_{k'}^{k''})^{a_{\hat{k}}}$ in $\mathcal{P}_{\mathcal{M},t}$ means that at epoch $t$ the sensor node $s_{k'}$ produces $a_{\hat{k}}$ data falling into bucket $k''$, contributing the bucket key $k_{k',t}^{k''}$ in total $a_{\hat{k}}$ times in $\mathcal{H}_{\mathcal{M},t}$. Here, the sensor node $s_{k'}$ producing the data falling into the bucket $\emptyset$ means that $s_{k'}$ senses nothing. Thus, we can infer the total amount of data falling into specified buckets at epoch $t$. Recall that the authority is aware of the topology of the aggregation tree. Thus, after the prime factor decomposition of $\mathcal{P}_{\mathcal{M},t}$, the authority can reconstruct $\mathcal{H}_{\mathcal{M},t}$ according to the derived $a_{\hat{k}}$'s and $\mathfrak{p}_{\hat{k}}$'s by its own effort, because it knows $K_{i,t}$ and $k_{i,t}^V$, $\forall i\in\{1,\dots,N-1,\mathcal{M}\}$, $\forall t\geq 0$, $\forall V\in\mathcal{V}$. Therefore, we know that the $\mathcal{H}_{\mathcal{M},t}$ reconstructed by the authority is equal to the received $\mathcal{H}_{\mathcal{M},t}$ if and only if the received $\mathcal{P}_{\mathcal{M},t}$ are considered authentic. When the verification of $\mathcal{P}_{\mathcal{M},t}$ fails, $\mathcal{M}$ is considered compromised. When the verification of $\mathcal{P}_{\mathcal{M},t}$ is successful, the authority decrypts all the received encryptions, and checks whether the number of query-results falling into the buckets in $\mathcal{A}$ matches those indicated by $\mathcal{P}_{\mathcal{M},t}$. If and only if there are matches in all the buckets in $\mathcal{A}$, the received query-results are considered complete.

\subsection{Securing Top-$k$ Queries (STQ)}\label{sec: Securing Top-$k$ Queries (STQ)}
Basically, the proposed STQ scheme for securing top-$k$ query is built upon SRQ in that both confidentiality-preserving reporting and query-result completeness verification phases in SRQ are exploited. In particular, based on the proof generated in SRQ, since it can know which buckets contain data, the authority can also utilize such information to examine the completeness of query-results of top-$k$ query. In other words, top-$k$ query can be secured by the use of the SRQ scheme. Because of the similarity between the SRQ and STQ schemes, some details of the STQ scheme will be omitted in the following description.

Here, a \emph{bucket data set} is defined to be composed of bucket IDs. We use a $d$-dimensional tuple, $(v_1,\dots,v_d)$, where $v_g\in[1,w]$, to represent the bucket IDs in a bucket data set. With this representation, we can use the ranking function, $R$, to calculate the ranking value of each bucket ID. Assume that the $\nu$-th interval in the $g$-th attribute contains the values in $[u_{g,\nu}^{\ell}, u_{g,\nu}^h]$. The ranking value of the bucket ID, $(v_1,\dots,v_d)$, is evaluated as $R(\frac{u_{1,v_1}^{\ell}+u_{1,v_1}^h}{2},\dots,\frac{u_{d,v_d}^{\ell}+u_{d,v_d}^h}{2})$, where the $d$-dimensional tuple, $(\frac{u_{1,v_1}^{\ell}+u_{1,v_1}^h}{2},\dots,\frac{u_{d,v_d}^{\ell}+u_{d,v_d}^h}{2})$, whose individual entry is simply averaged over the minimum and maximum values in each interval, acts as the representative of the bucket $(v_1,\dots,v_d)$ for simplicity.

Recall that we simply assume that the data with the first $k$ smallest ranking values are desired. The general form of the message sent from $s_i$ to its parent node at the end of epoch $t$ is $\langle i,t,\mathcal{E}_{i,t},\mathcal{B}_{i,t},\mathcal{H}_{i,t},\mathcal{P}_{i,t}\rangle$, where $\mathcal{E}_{i,t}$, $\mathcal{B}_{i,t}$, $\mathcal{H}_{i,t}$, and $\mathcal{P}_{i,t}$ are the same as those defined in SRQ. Assume that $\zeta_1,\dots,\zeta_k\in \mathcal{V}$ are $k$ bucket IDs in the bucket data set whose ranking values are among the first $k$ smallest ones. According to $\mathcal{B}_{\mathcal{M},t}$, $\mathcal{M}$ can calculate the ranking values of bucket IDs in $\mathcal{B}_{\mathcal{M},t}$ and, therefore, knows $\zeta_1,\dots,\zeta_k$. To answer a top-$k$ query, $\langle \mathcal{C}, t, k\rangle$, the storage node $\mathcal{M}$ reports the bucket IDs, $\zeta_1,\dots,\zeta_k$, and their corresponding encrypted data, along with $\mathcal{H}_{\mathcal{M},t}$ and $\mathcal{P}_{\mathcal{M},t}$, to the authority because it can be known that the data with the first $k$ smallest ranking values must be within $\zeta_1,\dots,\zeta_k$. After receiving the query-result, the authority can first verify the authenticity of $\mathcal{P}_{i,t}$ by using $\mathcal{H}_{i,t}$, and verify the query-result completeness by using $\mathcal{P}_{i,t}$. Note that both of the above verifications can be performed in a way similar to the one described in Sec. \ref{sec: Securing Range Queries (SRQ)}. Actually, after receiving $\mathcal{P}_{i,t}$, the authority knows which buckets contain data and the amount of data. Hence, knowing $u_{g,\nu}^{\ell}$ and $u_{g,\nu}^h$, $\forall g\in[1,d]$, $\forall \nu\in[1,w]$, the authority can also obtain $\zeta_1,\dots,\zeta_k$. Afterwards, what the authority should do is to check if it receives the bucket IDs, $\zeta_1,\dots,\zeta_k$, and if the number of data in bucket $\zeta_{g'}$, $g'\in[1,k]$, is consistent with the number indicated by $\mathcal{P}_{i,t}$. If and only if these two verifications pass, the authority considers the received query-result to be complete and extracts the top-$k$ result from the encrypted data sent from $\mathcal{M}$.

\subsection{Securing Skyline Queries (SSQ)}\label{sec: Securing Skyline Queries (SSQ)}
To support secure skyline query in sensor networks, in the following we first present a naive approach as baseline, and then propose an advanced approach that employs a grouping technique for simultaneously reducing the computation and communication cost.

\subsubsection{Baseline scheme}\label{sec: Baseline scheme}
To ensure the data confidentiality and authenticity, as in the SRQ and STQ schemes, the sensed data are also encrypted by using the bucket scheme mentioned in Sec. \ref{sec: Confidentiality-preserving reporting}. At the end of epoch $t$, each $s_i$ broadcasts its sensor ID, all the sensed data encrypted by key $K_{i,t}$, and the proper bucket IDs to all the nodes within the same cell. Then, according to the broadcast messages, each sensor node $s_i$ at epoch $t$ has a bucket data set composed of the bucket IDs extracted from broadcast messages and the bucket IDs corresponding to its own sensed data. In fact, the bucket data sets constructed by different nodes in the same cell at epoch $t$ will be the same. Treating these bucket IDs as data points, $s_i$ can find the set $\Phi_t$ of \emph{skyline buckets} that are defined as the bucket IDs not dominated by the other bucket IDs. Here, since the bucket IDs are represented also by $d$-dimensional tuples, the notion of domination is the same as the one defined in the query model of Sec. \ref{sec: System Model}. Define \emph{quasi-skyline data} as the set of data falling into the skyline buckets. It can be observed that the set of skyline data must be a subset of quasi-skyline data. After doing so, each node can locally find\footnote{The design of an algorithm for efficiently finding the skyline data given a data set is a research topic, but is beyond the scope of this paper. We consider the naive data-wise comparison-based algorithm with running time $O(n^2)$ if the size of the data set is $n$, but each node, in fact, can implement an arbitrary algorithm for finding skyline data in our setting.} the quasi-skyline data, although there could be the cases where superfluous data are also included. At the end of epoch $t$, if $K_{i,t}$ is smaller than a pre-determined threshold, then $s_i$ sends its sensor ID and $hash_{K_{i,t}}(\Phi_t)$ to $\mathcal{M}$. Here, $hash_{K_{i,t}}(\Phi_t)$ works as a kind of proof so that it can be used for checking the query-result completeness. Note that only $hash_{K_{i,t}}(\Phi_t)$ needs to be transmitted to $\mathcal{M}$ because $\mathcal{M}$ also receives all the encrypted data and bucket IDs, and can calculate the quasi-skyline data by itself after message broadcasting.

To answer the skyline query $\langle \mathcal{C}, t\rangle$ the storage node reports the quasi-skyline data calculated at epoch $t$ and the hash values received at epoch $t$ to the the authority. Since the authority knows the threshold and $K_{i,0}$ $\forall i\in[1,N]$, it will expect to receive hash values from a set of sensor nodes whose keys are smaller than the threshold. Unfortunately, due to the network-wide broadcast, this baseline scheme works but is inefficient in terms of communication overhead. Hence, an efficient SSQ scheme exploiting a grouping strategy is proposed as follows.

\subsubsection{Grouping technique}\label{sec: Grouping technique}
Like the baseline scheme, the bucket scheme mentioned in Sec. \ref{sec: Confidentiality-preserving reporting} is also used. Given a data set $\mathcal{O}$ and a collection $\{\mathcal{O}_1, \cdots, \mathcal{O}_{\tau}\}$ of subsets of $\mathcal{O}$ satisfying $\bigcup_{j=1}^{\tau}\mathcal{O}_j=\mathcal{O}$ and $\mathcal{O}_j\cap \mathcal{O}_{j'}=\emptyset$, $\forall j\neq j'$ for $\tau\geq 1$. An observation is that the skyline data of $\mathcal{O}$ must be a subset of the union of the skyline data of $\mathcal{O}_j$, $\forall j\in[1,\tau]$. Thus, the key idea of our proposed SSQ scheme is to partition the sensor nodes in a cell into groups so that broadcasting can be limited within a group, resulting in reduced computation and communication costs. In what follows, the SSQ scheme will be described in more detail.

The sensor nodes in a cell are divided into $\mu$ disjoint groups, $G_{\eta}$, $\forall \eta\in[1,\mu]$, each of which is composed of $|G_{\eta}|$ sensor nodes. The grouping needs to be performed only once right after the sensor deployment. Note that each group is formed by nearby sensor nodes and the grouping procedure is independent of the structure of the aggregation tree without affecting SRQ and STQ. Let \emph{cell-region} be part of the sensing region monitored by one specified cell. For example, with the assumption that the shape of each cell-region is approximately a square, as shown in Fig. \ref{fig: tiered_network}, and sensor nodes with uniform deployment are considered, the grouping can be achieved by simply dividing a cell-region into $\mu$ ($=\sqrt{N}$) sub-cell-regions. The sensor nodes in the same sub-cell-region form a group. Note that the square cell-region is assumed here for ease of explanation, but is not necessary for the grouping procedures\footnote{For example, the authority knowing the position of each node or the use of clustering algorithms can also divide nodes into groups. In general, after the localization, each sensor node can join the proper group possibly according to its geographic position when the grouping information such as the sizes of sensing region and cell-region are preloaded in sensor nodes.}.

At the end of epoch $t$, each sensor node $s_i$ broadcasts its sensor ID, its \emph{order seed} $\theta_{i,t}=hash_{K_{i,t}}(i)$, all the sensed data encrypted with the key $K_{i,t}$, and the proper bucket IDs to all the nodes within the same group. After doing so, as in baseline scheme, the sensor nodes in the same group can locally find the quasi-skyline data from the bucket data set whose entries are generated by the sensor nodes in the same group. Let $\Phi_{\eta,t}$ be the set of skyline bucket IDs and $\phi_{\eta,t}$ their corresponding quasi-skyline data encrypted by the sensor nodes in group $\eta$ using proper keys at epoch $t$. At the end of epoch $t$, if $\theta_{i,t}$ is among the first $\xi_1$ smallest ones in the set of order seeds in group $\eta$ at epoch $t$, where $\xi_1$ is a pre-determined threshold known by each node, then $s_i$ reports $\Phi_{\eta,t}$, $\phi_{\eta,t}$, its \emph{verification seed} $hash_{K_{i,t}}(\Phi_{\eta,t})$, and the IDs of sensor nodes generating $\phi_{\eta,t}$ to $\mathcal{M}$. In fact, $\xi_1=1$ is sufficient for the verification purpose. $\xi_1$, however, is also related to the resiliency against sensor node compromises. Thus, we still keep $\xi_1$ as a variable and defer the explanation of the purpose of $\xi_1$ to Sec. \ref{sec: Impact of Sensor Node Compromise}. Here, the purpose of verification seeds is that the completeness of quasi-skyline data can be guaranteed by exactly $\xi_1$ sensor nodes for each group, while the purpose of order seed is to guarantee that at each epoch exactly $\xi_1$ sensor nodes will send the verification seeds as the proofs.

To answer a skyline query, $\langle \mathcal{C}, t\rangle$, the storage node $\mathcal{M}$ reports a hash of all the received verification seeds, $h_t=hash(||_{i\in\Gamma_t}hash_{K_{i,t}}(\Phi_{i,t}))$, where $||$ denotes the bit-string concatenation and $\Gamma_t$ is the set of sensor nodes responsible for sending a hash value to $\mathcal{M}$ in each group at epoch $t$, the set of skyline bucket IDs, and their corresponding encrypted data received at epoch $t$ to the authority. Since it knows the threshold $\xi_1$ and $K_{i,t}$, $\forall i\in\{1,\dots,N-1,\mathcal{M}\}$, the authority will expect to receive a particular hash value from $\mathcal{M}$. If and only if the hash sent from the $\mathcal{M}$ matches the hash of the verification seeds calculated according to the knowledge of $\Gamma_t$ and $K_{i,t}$ by the authority itself, the received data are considered complete, and contain the skyline data.

\section{Performance Evaluation}\label{sec: Performance Evaluation}
We will focus on analyzing the critical issue of detecting an incomplete query-result in tiered networks. In this section, the detection probability and communication cost of query-result completeness verification in the proposed schemes will be analyzed. It is assumed that the number of hops between $\mathcal{M}$ and each sensor node is $\sqrt{n}$ for a collection of $n$ uniformly deployed nodes \cite{cp05}. In this section, both detection probability and communication cost are discussed at a fixed epoch $t$.

As the communication cost of encoding approach \cite{sl08} grows exponentially with the number of attributes, and some crosscheck approaches \cite{szz09,zsz09} have relatively low detection probability, in the following, we compare SRQ with only \emph{hybrid crosscheck} \cite{zsz09}, which achieves the best balance between the detection probability and communication cost in the literature. Note that, the parameter setting required in hybrid crosscheck is the same as that listed in \cite{zsz09}.

\subsection{Detection Probability}\label{sec: Detection Probability}
The detection probability is defined as the probability that the compromised storage node $\mathcal{M}$ is detected if it returns an incomplete query-result. With the fact that the larger the portion of query-result $\mathcal{M}$ drops, higher the probability that the authority detects it, we consider the worst case that only one bucket and its corresponding data in the query-result are dropped by $\mathcal{M}$ and the number of data sensed by a node is either $0$ or $1$ as the lower bound of detection probability.

\subsubsection{Detection probability for SRQ and STQ}\label{sec: Detection probability of SRQ and STQ}
\begin{figure}[h]
\centering
\subfloat[]{\label{fig: detection probability Y100}\includegraphics[width=0.26\textwidth]{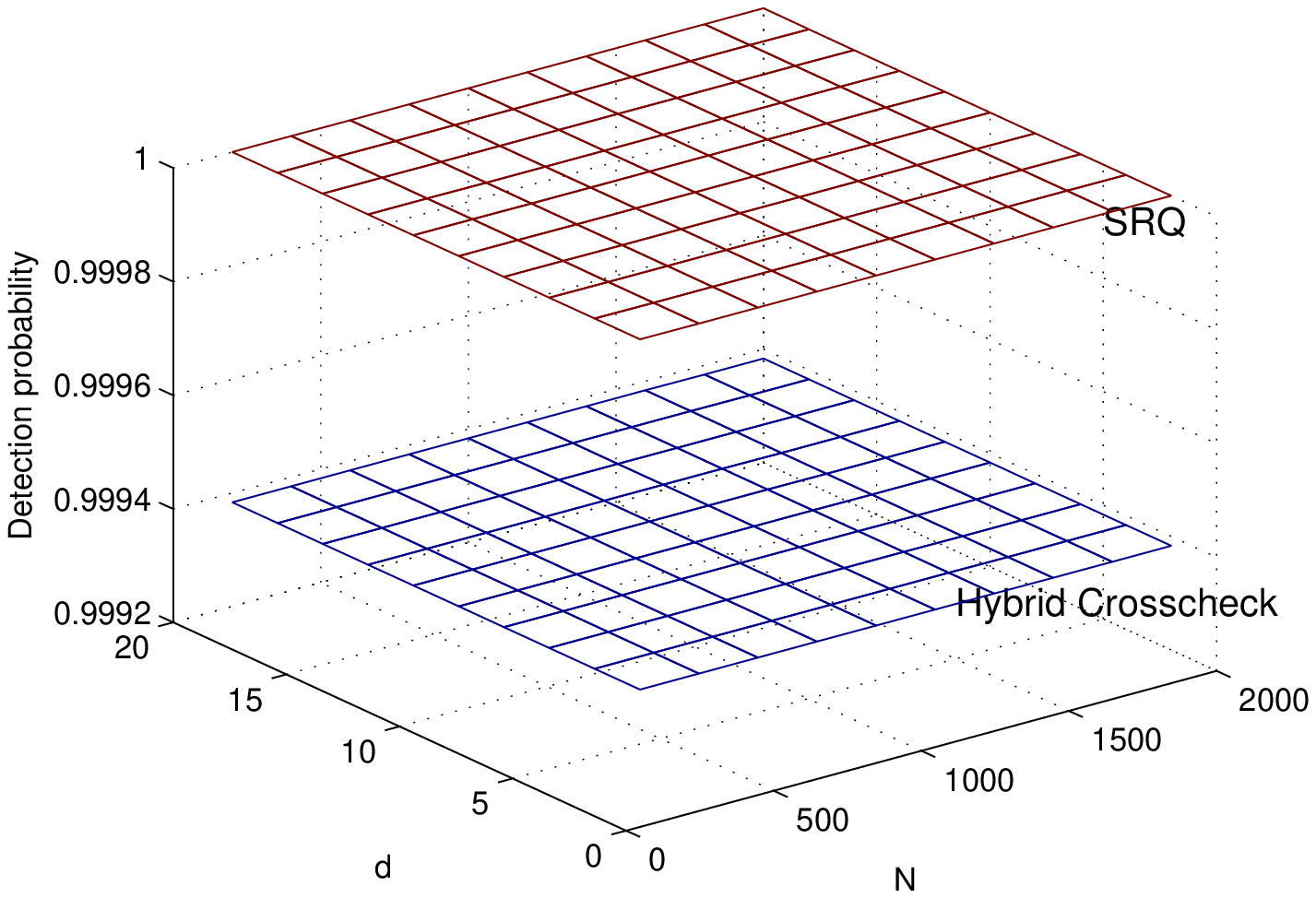}}
\subfloat[]{\label{fig: detection probability Y5000}\includegraphics[width=0.26\textwidth]{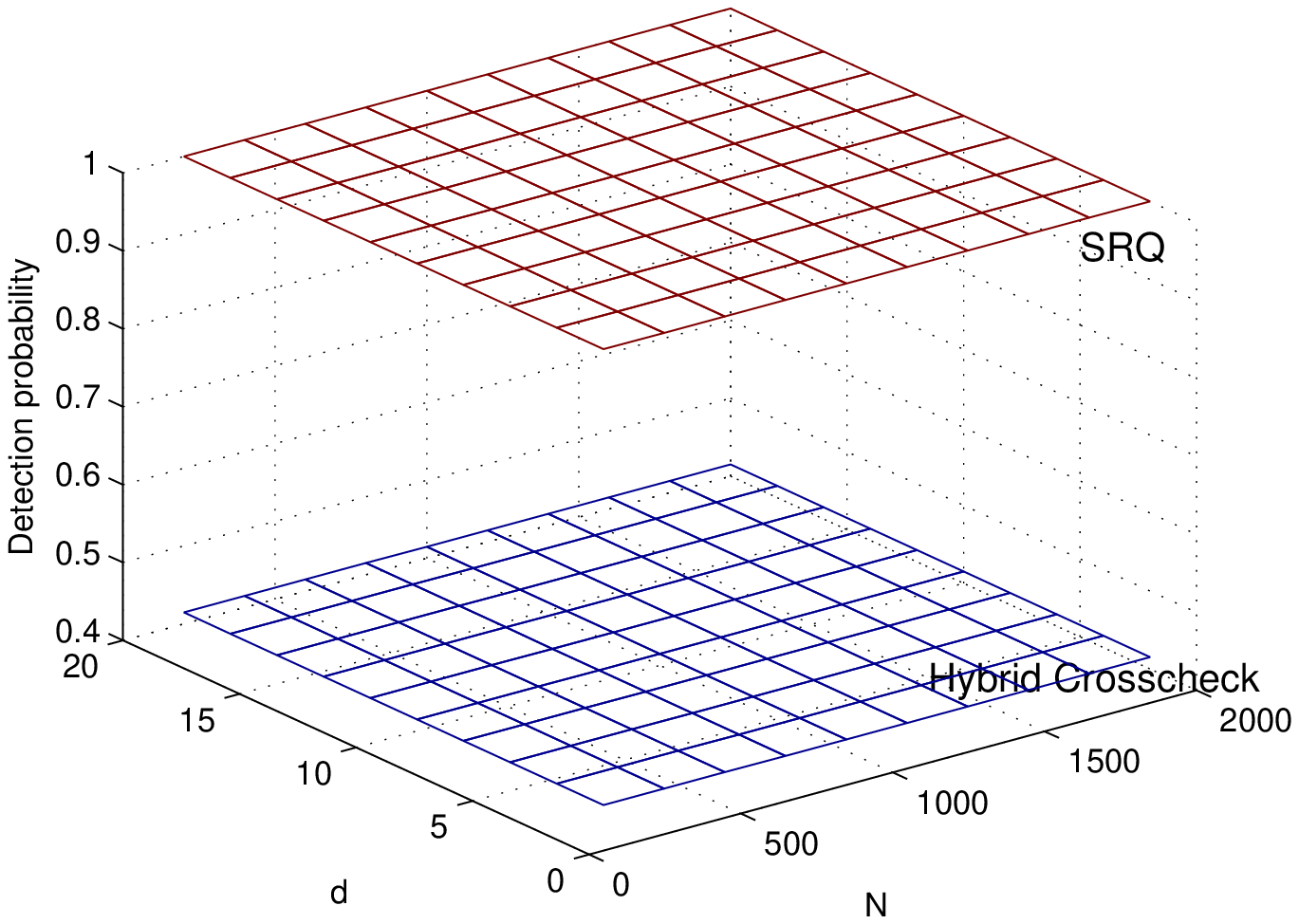}}
\caption{\scriptsize The detection probability of SRQ and hybrid crosscheck in the cases that (a) $Y=100$ and (b) $Y=5000$.} \label{fig: detection probability}
\end{figure}

To return an incomplete query-result without being detected by the authority, $\mathcal{M}$ should create a proof, $\langle\widehat{\mathcal{H}}_{\mathcal{M},t},\widehat{\mathcal{P}}_{\mathcal{M},t}\rangle$, corresponding to the incomplete query-result. Since bucket primes can be known by the adversary, $\widehat{\mathcal{P}}_{\mathcal{M},t}$ can be easily constructed. Nevertheless, $\widehat{\mathcal{H}}_{\mathcal{M},t}$ cannot be constructed, since the bucket keys of sensor nodes generating the bucket dropped by $\mathcal{M}$ are not known by the adversary. Therefore, only two options can be chosen by the adversary. First, the adversary can directly guess to obtain $\widehat{\mathcal{H}}_{\mathcal{M},t}$, with probability being $2^{-\ell_h}$, where $\ell_h$ is the number of bits output by a keyed hash function. This implies that the detection probability $\mathbb{P}_{det,1}^{SRQ}$ is $1-2^{-\ell_h}$ for the first case. Second, knowing the aggregation tree topology, the adversary can also follow the rule of SRQ to construct the $\widehat{\mathcal{H}}_{\mathcal{M},t}$ without considering the bucket key of dropped bucket. Assume that the probability that a sensor node has sensed data is $\delta$. The size of the bucket key pool can be derived as $(N-2)(2\delta+1-\delta)+2=N\delta+N-2\delta$. Thus, the probability for the adversary to guess successfully is $2^{-\ell_k(N\delta+N-2\delta)}$, where $\ell_k$ is the number of bits of a key, leading to the detection probability $\mathbb{P}_{det,2}^{SRQ}$ is $1-2^{-\ell_k(N\delta+N-2\delta)}$ for the second case. Overall, the final detection probability, $\mathbb{P}_{det}^{SRQ}$, is $\min\{\mathbb{P}_{det,1}^{SRQ},\mathbb{P}_{det,2}^{SRQ}\}$. On the other hand, as stated in Sec. \ref{sec: Securing Top-$k$ Queries (STQ)}, the STQ scheme is built upon the SRQ scheme. Thus, the detection probability, $\mathbb{P}_{det}^{STQ}$, will be the same as $\mathbb{P}_{det}^{SRQ}$. As Fig. \ref{fig: detection probability} depicts, the detection probability of SRQ is close to $1$ in any case. However, hybrid crosscheck is effective only when a few sensed data are generated in the network. Such a performance difference can be attributed to the fact that the sensed data in the network are securely and deterministically summarized in the proof of SRQ but they are probabilistically summarized in hybrid crosscheck.

\subsubsection{Detection probability for SSQ}\label{sec: Detection probability of SSQ}
To return an incomplete query-result without being detected by the authority, $\mathcal{M}$ should forge a proof, \emph{i.e.}, a hash $\widehat{h}_t$ of all the received verification seeds, corresponding to the incomplete query-result. Therefore, only two options can be chosen by the adversary. First, the adversary can directly guess a hash value $\widehat{h}_t$. The probability for the adversary to guess successfully is $2^{-\ell_h}$, implying the detection probability $\mathbb{P}_{det,1}^{SSQ}$ is $1-2^{-\ell_h}$ in this case. Second, the adversary can also follow the rule of SRQ to construct $\widehat{h}_t$ for incomplete data. In this case, the adversary is forced to guess $\xi_1$ keys for each one of $\mu$ groups, leading to the probability of success guess being $2^{-\ell_k \mu \xi_1}$ and the detection probability being $\mathbb{P}_{det,2}^{SSQ}=1-2^{-\ell_k \mu \xi_1}$. The final detection probability, $\mathbb{P}_{det}^{SSQ}$, is thus, $\min\{\mathbb{P}_{det,1}^{SSQ},\mathbb{P}_{det,2}^{SSQ}\}$. Obviously, $\mathbb{P}_{det}^{SSQ}$ will also be close to $1$ when appropriate key length or hash function is selected.

\subsection{Communication Cost}\label{sec: Communication Cost}
The communication cost, $\mathbb{T}$, is defined as the number of bits in the communications required for the proposed schemes. We are mainly interested in the asymptotic result in terms of $d$ and $N$ because they reflect the scalability of the number of attributes and the network size, respectively. We do not count the number of bits in representing data, $\mathcal{E}_{i,t}$, since the sending of $\mathcal{E}_{i,t}$ is necessary in any data collection scheme. We further assume that there are on average $p_{dtob}Y$ data buckets, where $0\leq p_{dtob}\leq 1$, generated in cell $\mathcal{C}$.

\subsubsection{Communication Cost of SRQ and STQ} Each sensor node $s_i$ in SRQ is required to send $\mathcal{B}_{i,t},\mathcal{H}_{i,t}$, and $\mathcal{P}_{i,t}$ to its parent node. Nevertheless, $\mathcal{B}_{i,t},\mathcal{H}_{i,t}$, and $\mathcal{P}_{i,t}$ can be aggregated along the path in the aggregation tree. As a consequence, $s_i$ actually has only one-hop broadcast containing $\mathcal{B}_{i,t},\mathcal{H}_{i,t}$, and $\mathcal{P}_{i,t}$ once at each epoch. In addition, to answer a range query, $\mathcal{M}$ is responsible for sending $\mathcal{H}_{\mathcal{M},t}$, $\mathcal{P}_{\mathcal{M},t}$, and the bucket IDs in $\mathcal{A}$. In summary, the communication cost, $\mathbb{T}^{SRQ}$, can be calculated as $(N-1)(\ell_h+\ell_P)+p_{dtob}Y\lceil\log w\rceil d\log N+\ell_h+\ell_P+|\mathcal{A}|\lceil\log w\rceil d=O(N+d\log N)$, where $\ell_P$ is the number of bits used to represent the bucket prime $\mathcal{P}_{\mathcal{M},t}$. Due to the similarity between SRQ and STQ, the communication cost, $\mathbb{T}^{STQ}$, can also be calculated as $O(N+d\log N)$ in a way similar to the one for obtaining $\mathbb{T}^{SRQ}$.

\begin{figure}[h]
\centering
\subfloat[]{\label{fig: communication cost Y100}\includegraphics[width=0.26\textwidth]{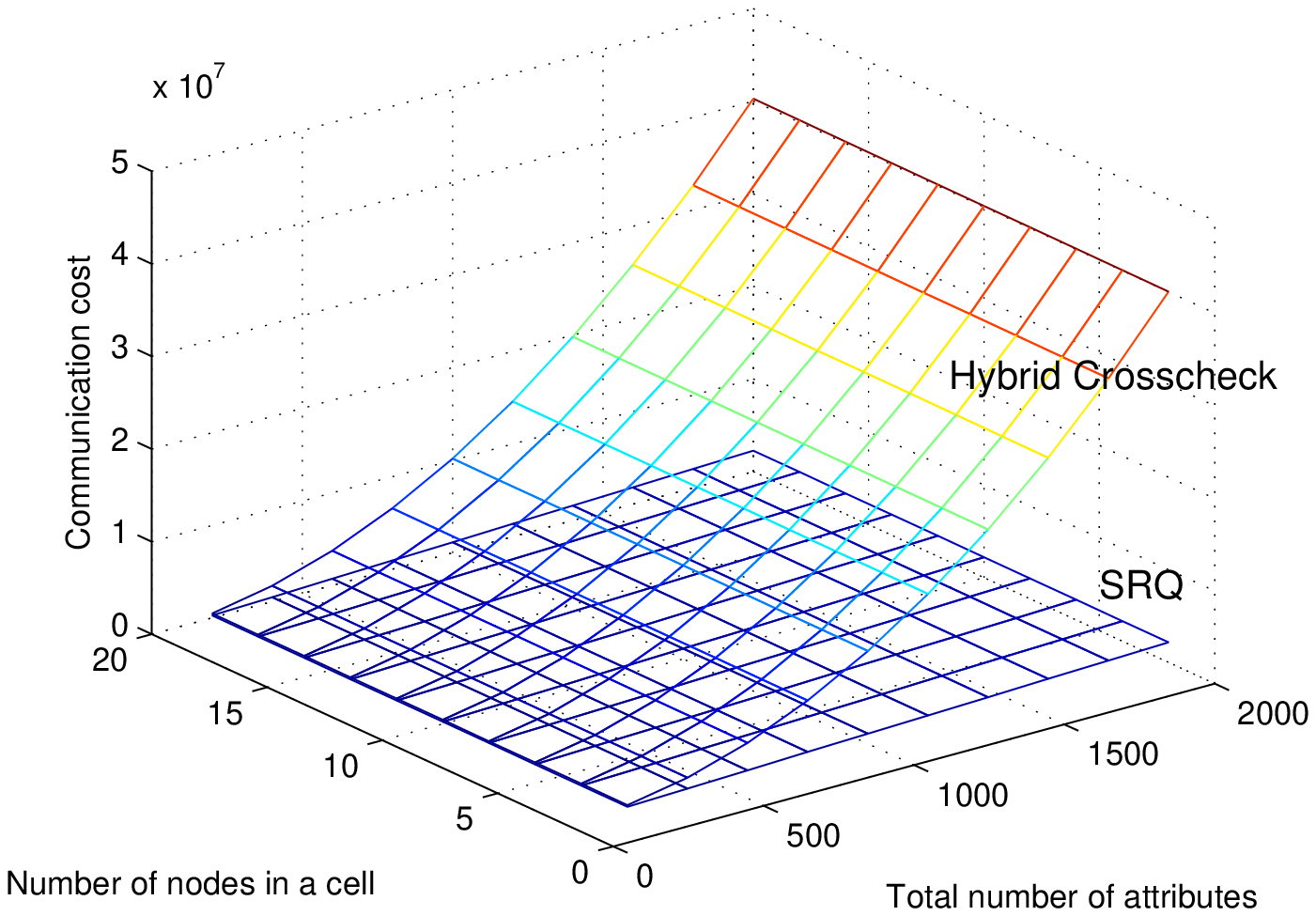}}
\subfloat[]{\label{fig: communication cost Y5000}\includegraphics[width=0.26\textwidth]{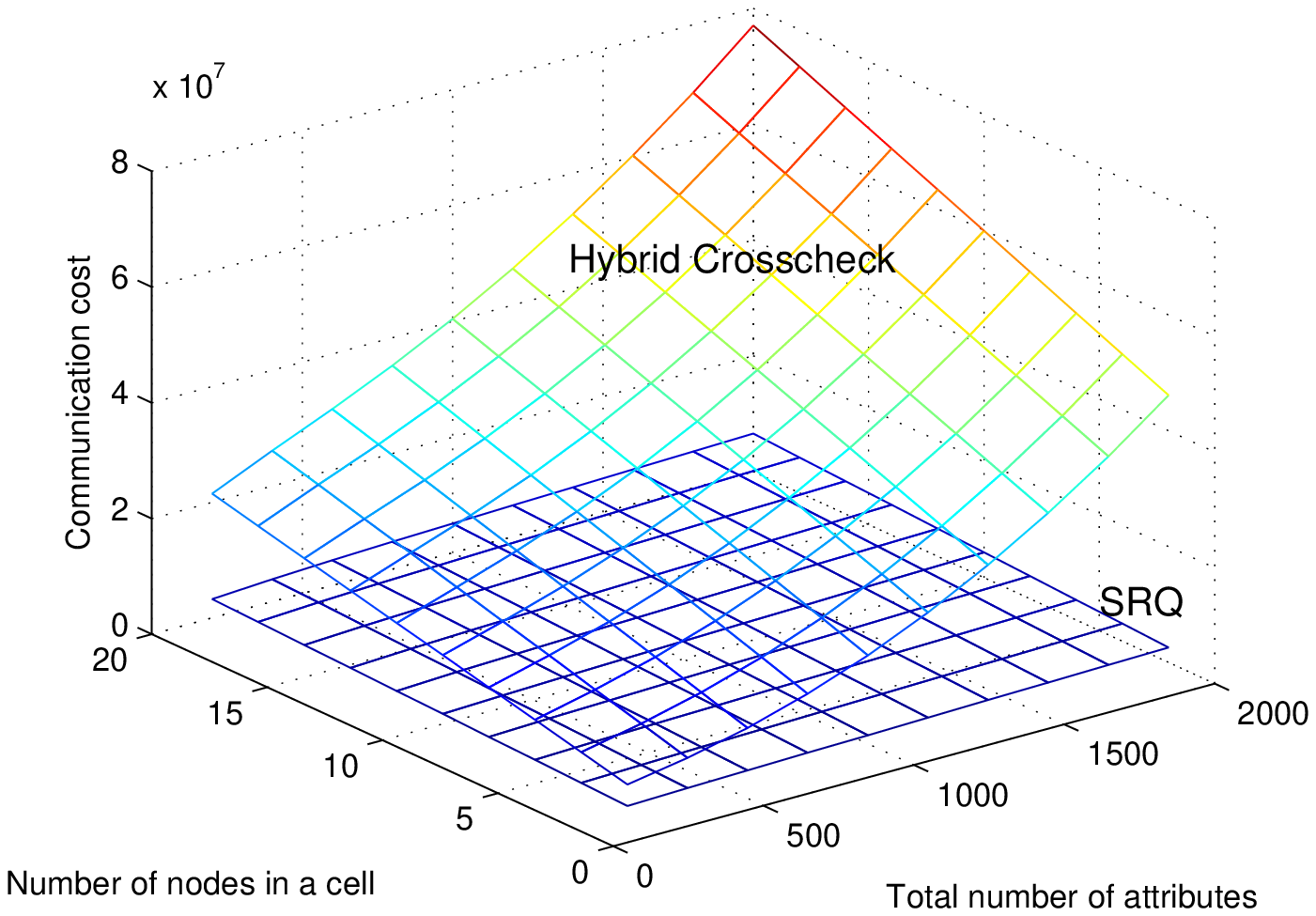}}
\caption{\scriptsize The communication cost of SRQ and hybrid crosscheck in the cases that (a) $Y=100$ and (b) $Y=5000$.} \label{fig: communication cost}
\end{figure}

As shown in Fig. \ref{fig: communication cost} where the parameters $\ell_h=80$ and $\ell_P=1000$ are used, the communication cost of SRQ is significantly lower than that of hybrid crosscheck. More specifically, as the communication cost of hybrid crosscheck can be asymptotically represented as $O(N^2+Nd)$ and will be drastically increased with $N$ and $d$, the proposed SRQ scheme, however, exhibits low communication cost regardless of the amount of sensed data in the network due to the fact that the size of the proof used in SRQ is always a constant. Hence, the communication cost of SRQ will be dominated by the aggregation procedure, the average hop distance between $\mathcal{M}$ and each node, and the transmission of bucket IDs, resulting $O(N+d\log N)$ communication cost.

\subsubsection{Communication Cost of SSQ}
\begin{figure}[h]
\centering
\subfloat[]{\label{fig: SSQ communication cost Y100}\includegraphics[width=0.26\textwidth]{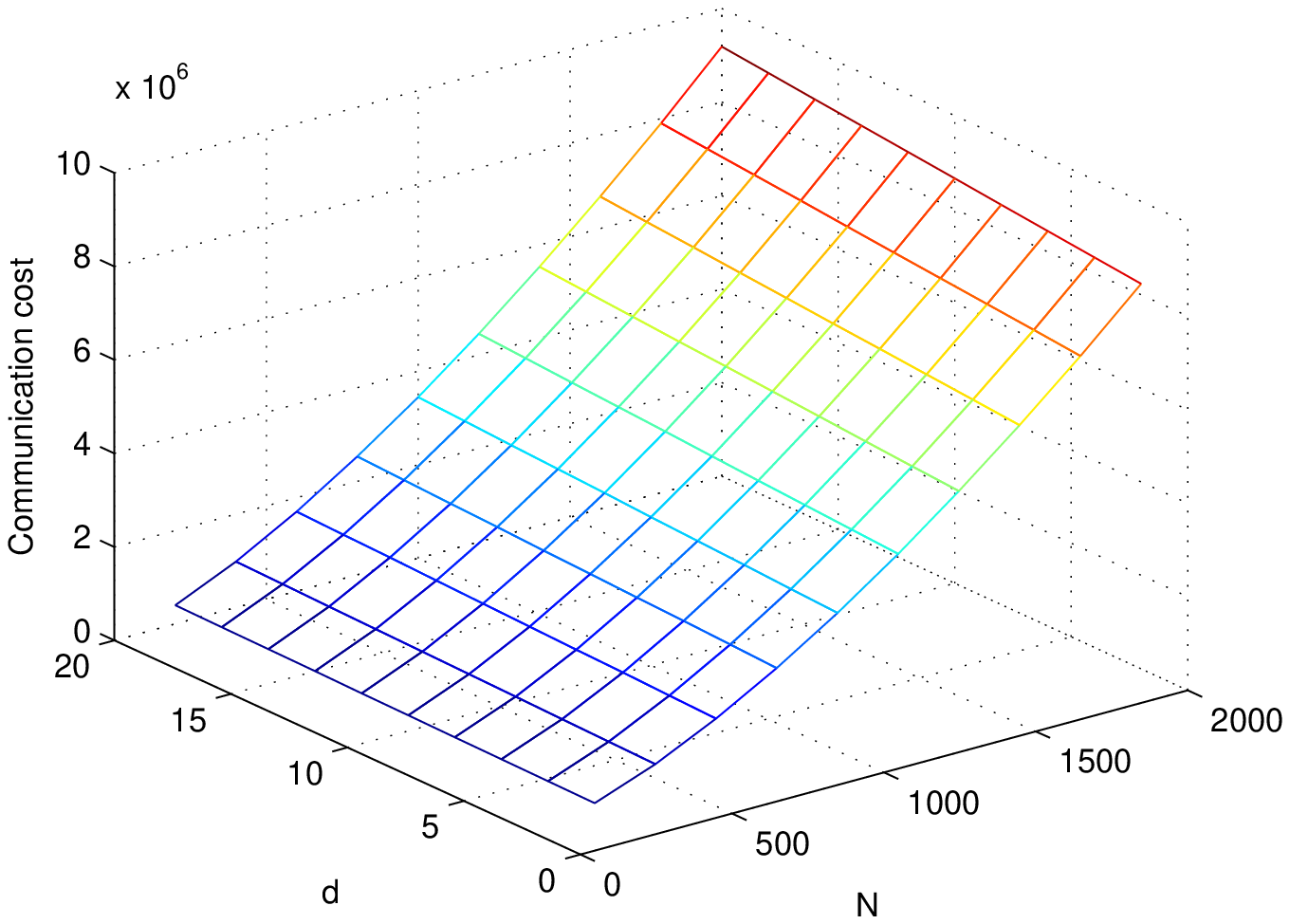}}
\subfloat[]{\label{fig: SSQ communication cost Y5000}\includegraphics[width=0.26\textwidth]{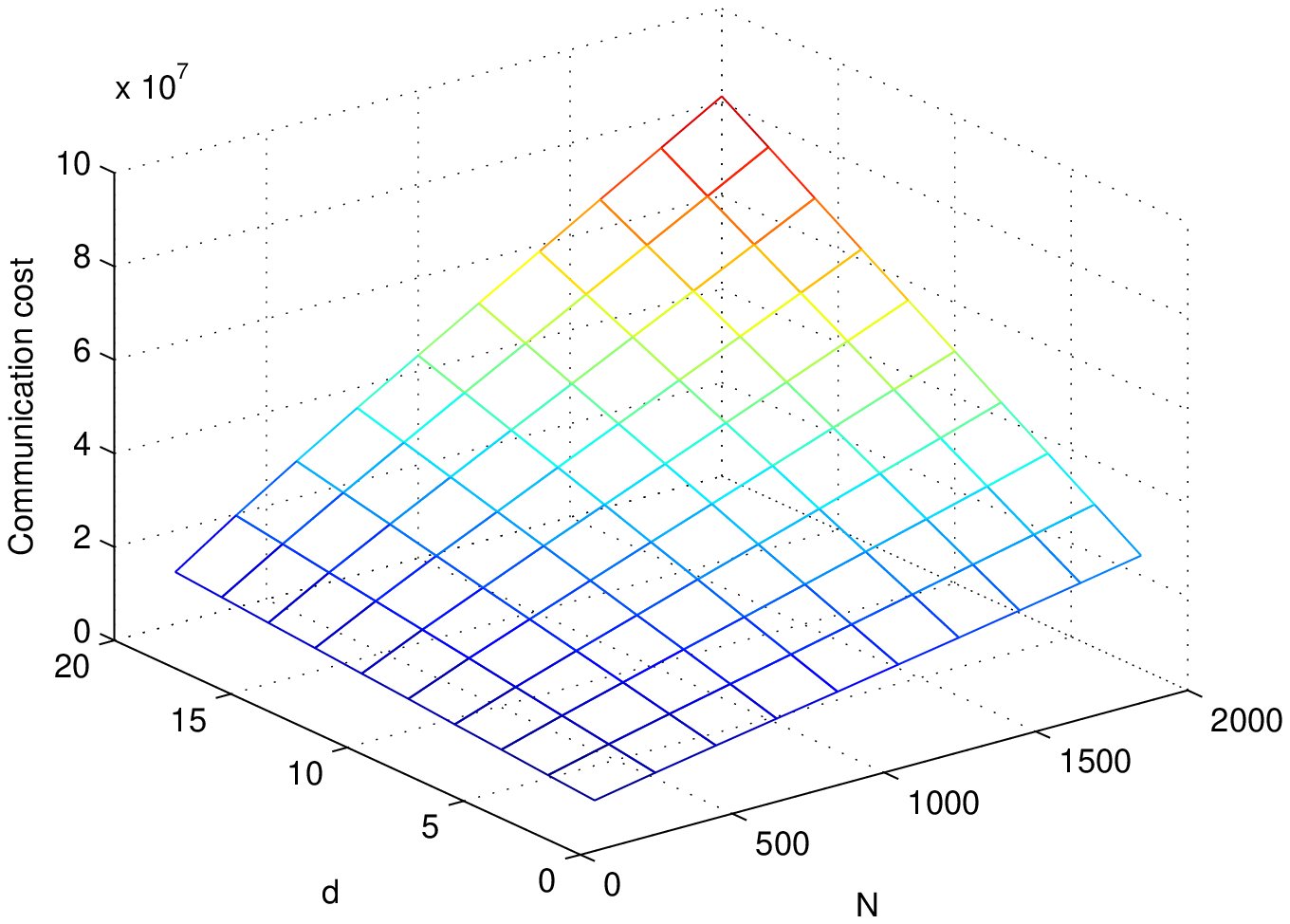}}
\caption{\scriptsize The communication cost of SSQ for (a) $Y=100$ and (b) $Y=5000$.} \label{fig: SSQ communication cost}
\end{figure}

Since grouping is only performed once after sensor deployment, we ignore its communication cost. Note that, in the following, the communication cost of the data should be counted because it is involved in the design of SSQ. After the grouping, each $s_i$ broadcasts the data bucket IDs, sensor ID, and the order seed to all the nodes within the same group. Assuming that the nodes employ a duplicate suppression algorithm, by which each node only broadcasts a given message once, one node should broadcast a message with $\frac{Y}{N}\ell_d+\frac{p_{dtob}Y}{N}\lceil\log w\rceil d+\ell_{id}+\ell_h$ bits, where $\ell_d$ is the average size of a datum and $\ell_{id}$ is the number of bits required to represent sensor IDs, resulting in communication cost of $C_1=|G_\eta|^2(\frac{Y}{N}\ell_d+\frac{p_{dtob}Y}{N}\lceil\log w\rceil d+\ell_{id}+\ell_h)$ bits in each group. Let $0\leq p_q\leq 1$ be the average ratio of the quasi-skyline data to all the sensed data. $|\Phi_{\eta,t}|$ is equal to $p_{dtob}p_qY$ on average. Then, once the the order seed is among the first $\xi_1$ smallest ones of the order seeds in a group, $s_i\in G_\eta$ is required for sending $\Phi_{\eta,t}$, $\phi_{\eta,t}$, $hash_{K_{i,t}}(\Phi_{\eta,t})$, its sensor ID, and the IDs of sensor nodes generating buckets in $\Phi_{\eta,t}$ to $\mathcal{M}$, implying the communication cost of $C_2=p_{dtob}p_qY\lceil\log w\rceil d+p_qY\ell_{d}+\ell_h+\ell_{id}+|G_{\eta}|\ell_{id}$ bits in the worst case. Note that the verification seeds cannot be aggregated, although the verification seed can also be delivered along the path to $\mathcal{M}$ on the aggregation tree. $\mathcal{M}$ needs to send the skyline bucket IDs, its sensor ID, and a hash $h_t$ as the proof to the authority for answering the query. The communication cost of $\mathcal{M}$ is $C_3=\ell_{id}+\ell_h+\mu\sum_{\eta=1}^{\mu}p_{dtob}p_qY\lceil\log w\rceil d+p_qY\ell_d$. Consequently, the upper bound of communication cost, $\mathbb{T}^{SSQ}$, can be obtained as $\mu C_1+\mu\xi_1C_2\log N+C_3=O(N^{\frac{3}{2}}+Nd)$ when $\mu=\sqrt{N}$ and $|G_\eta|=\sqrt{N}$, $\forall\eta\in[1,\mu]$. By similar derivation, the communication cost of the baseline scheme is $O(N^2d)$. Thus, exploiting the proposed grouping technique does reduce the required communication cost. The trends of communication cost in SSQ are shown in Fig. \ref{fig: SSQ communication cost}.

\section{Impact of Sensor Node Compromise}\label{sec: Impact of Sensor Node Compromise}
In the previous discussions, we have ignored the impact of the compromise of sensor nodes. In practice, the adversary could take the control of sensor nodes to enhance the ability of performing malicious operations. The notation $\mathfrak{s}$ is used to denote a set of random sensor nodes compromised by the adversary. In \emph{collusion attack} considered here, $\mathcal{M}$ colludes with $\mathfrak{s}$ in the hope that more portions of query-results generated by innocent sensor nodes can be dropped. Since crosscheck approaches \cite{zsz09} suffer from collusion attacks, the impact of collusion attack on secure range query for tiered sensor networks was addressed in \cite{zsz09}. Their proposed method is \emph{random probing}, by which the authority occasionally checks if there is no data sensed by some randomly selected sensor nodes by directly communicating with them. Random probing, however, can only discover the incomplete query-results with an inefficient but possibly lucky way and cannot identify $\mathfrak{s}$.

On the other hand, in this paper, we identify a new Denial-of-Service attack never addressed in the literature, called \emph{false-incrimination attack}, by which $\mathfrak{s}$ provides false sub-proofs to the innocent storage node so that the innocent storage node will be regarded as the compromised one and be revoked. Unfortunately, all the prior works \cite{sl08,szz09,zsz09} suffer from this attack. In summary, with minor modifications involved, our proposed SRQ, STQ, and SSQ schemes are resilient against both the collusion attack and false-incrimination attack.

It should be especially noted that, in the following discussion of SRQ, we temporarily make two unrealistic assumptions that the compromised nodes can only disobey the procedures of the proposed schemes\footnote{In other words, compromised nodes are assumed to \emph{not} inject bogus sensor readings. They can only manipulate its own subproof and the proofs sent from its descendant sensor nodes.} and the (subtree) proofs (defined later) will not be manipulated by the compromised storage and sensor nodes, in order to emphasize on the effectiveness of our proposed technique in identifying compromised nodes. Nevertheless, these two assumptions will be relaxed later.

\textbf{Impact of Sensor Node Compromise on SRQ and STQ.} Under the above two assumptions, the SRQ scheme is inherently resilient against collusion attack, because, regardless of the existence and position of $\mathfrak{s}$, the proper bucket keys will be embedded into the proofs and cannot be removed by $\mathfrak{s}$. Nevertheless, SRQ could be vulnerable to the false-incrimination attack since false sub-proofs injected by $\mathfrak{s}$ will be integrated with the other correct sub-proofs to construct a false proof, leading to the revocation of innocent $\mathcal{M}$. Here, we present a novel technique called \emph{subtree sampling} enabling SRQ, with a slight modification, to efficiently mitigate the threat of false-incrimination attacks. The idea of \emph{subtree sampling} is to check if the proof constructed by the nodes in a random subtree with fixed depth is authentic so as to perform the attestation only on the remaining suspicious nodes. Let $m$ be a user-selected constant indicating the subtree depth. In the modified SRQ scheme, once receiving $\langle j_\rho, t, \mathcal{E}_{j_\rho,t}, \mathcal{B}_{j_\rho,t}, \mathcal{H}_{j_\rho,t}, \mathcal{P}_{j_\rho,t}, \vartheta_{j_\rho,t}^{0},\dots,\vartheta_{j_\rho,t}^{m-1}\rangle$, $j_{\rho}\in[1,N-1]$, $\forall \rho\in [1,\chi]$, from its $\chi$ children, $s_{j_1},\dots,s_{j_\chi}$, each $s_{i}$ calculates $\mathcal{E}_{j_\rho,t}$, $\mathcal{B}_{j_\rho,t}$, $\mathcal{H}_{j_\rho,t}$, and $\mathcal{P}_{j_\rho,t}$ as in the original SRQ scheme. Note that, if $s_i$ is a leaf node on the aggregation tree, it is assumed that $s_i$ receives $\langle\emptyset,\emptyset,\emptyset,\emptyset,0,1,\emptyset,\dots,\emptyset\rangle$. In the modified SRQ scheme, however, $s_i$ additionally performs the following operations. Assume that $\bar{H}_{j_\rho,t}^\upsilon$, $\bar{P}_{j_\rho,t}^{\upsilon}\in\vartheta_{j_\rho,t}^\upsilon$, $\forall \upsilon\in[0,m-1]$, $\forall j_\rho\in[1,N-1]$, and $\forall \rho\in[1,\chi]$. $s_i$ calculates $\bar{H}_{i,t}^\upsilon=hash_{K_{i,t}}(\sum_{\rho=1}^{\chi}\bar{H}_{j_{\rho},t}^{\upsilon-1}+H_{i,t})$ and $\bar{P}_{i,t}^\upsilon=\prod_{\rho=1}^{\chi}\bar{P}_{j_{\rho},t}^{\upsilon-1}\cdot P_{i,t}$, $\forall \upsilon\in[1,m]$. $s_i$ also calculates $\bar{H}_{i,t}^0=H_{i,t}$ and $\bar{P}_{i,t}^0=P_{i,t}$, where $H_{i,t}$ and $P_{i,t}$ are computed in a way stated in Sec. \ref{sec: Securing Range Queries (SRQ)}. Then, $\bar{H}_{i,t}^\upsilon$ and $\bar{P}_{i,t}^\upsilon$ are assigned to set $\vartheta_{i,t}^{\upsilon}$, $\forall\upsilon\in[0,m-1]$. If $hash_{K_{i,t}}(i)\leq\xi_2$, where $\xi_2$ is a pre-determined threshold known by each node and will be analyzed later, then $s_i$ sends $\vartheta_{i,t}^m$ to (possibly compromised) $\mathcal{M}$. Let $T_i^m$ be a subtree of the underlying aggregation tree, rooted at $s_i$ with depth $m$. $\vartheta_{i,t}^m$ generated by $s_i$ can be thought of as the \emph{subtree proof} of the data sensed by the nodes in $T_i^m$. Finally, $s_i$ sends $\langle i, t, \mathcal{E}_{i,t}, \mathcal{B}_{i,t}, \mathcal{H}_{i,t}, \mathcal{P}_{i,t}, \vartheta_{i,t}^{0},\dots,\vartheta_{i,t}^{m-1}\rangle$ to its parent node. Let $W_t$ be the \emph{witness set} of sensor nodes satisfying $hash_{K_{i,t}}(i)\leq \xi_2$ at epoch $t$. The nodes in $W_t$ are called \emph{witness nodes} at epoch $t$.

\begin{figure}[h]
\centering
\subfloat[]{\label{fig: diagram goodcase}\includegraphics[width=0.25\textwidth]{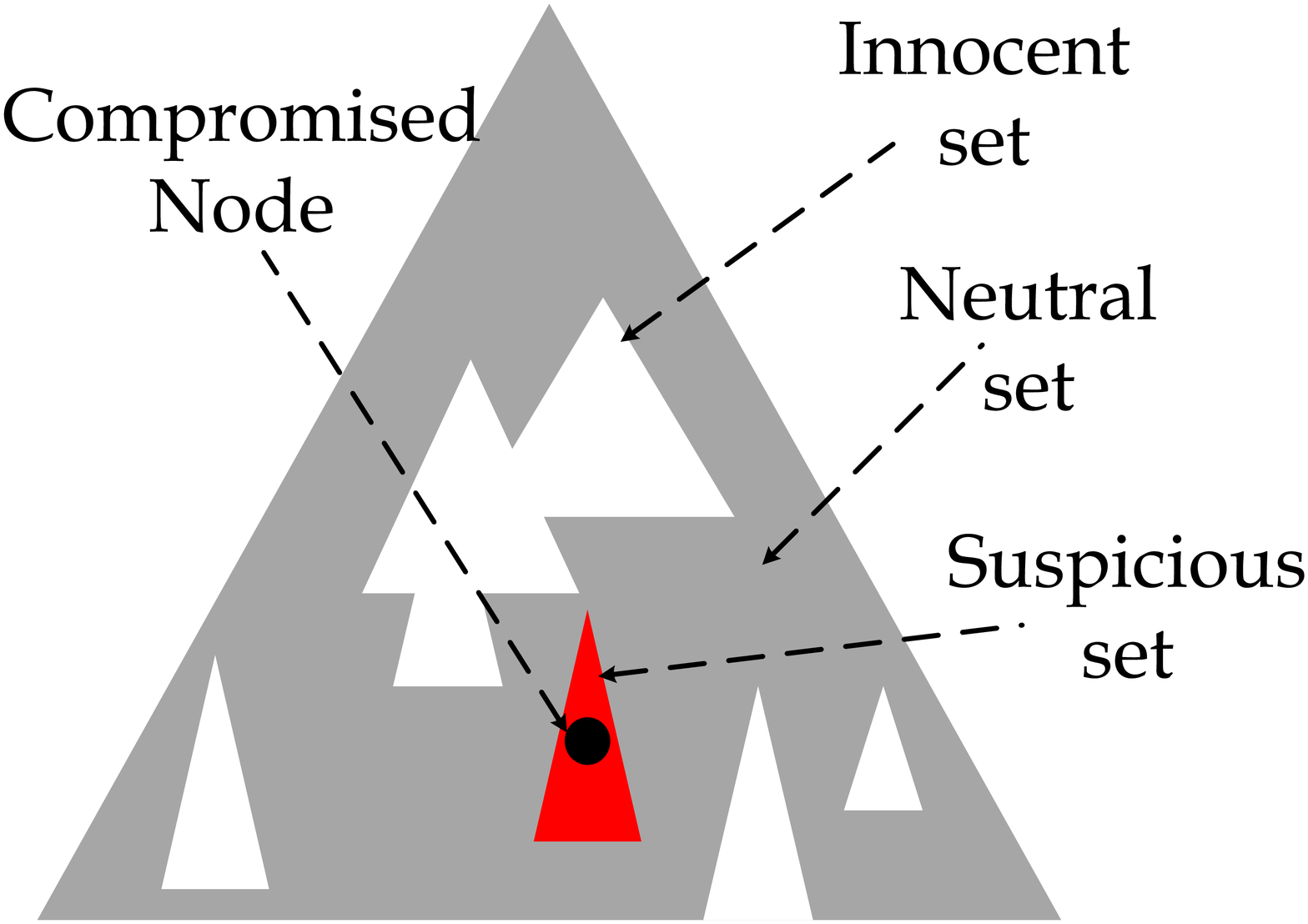}}
\subfloat[]{\label{fig: diagram badcase}\includegraphics[width=0.25\textwidth]{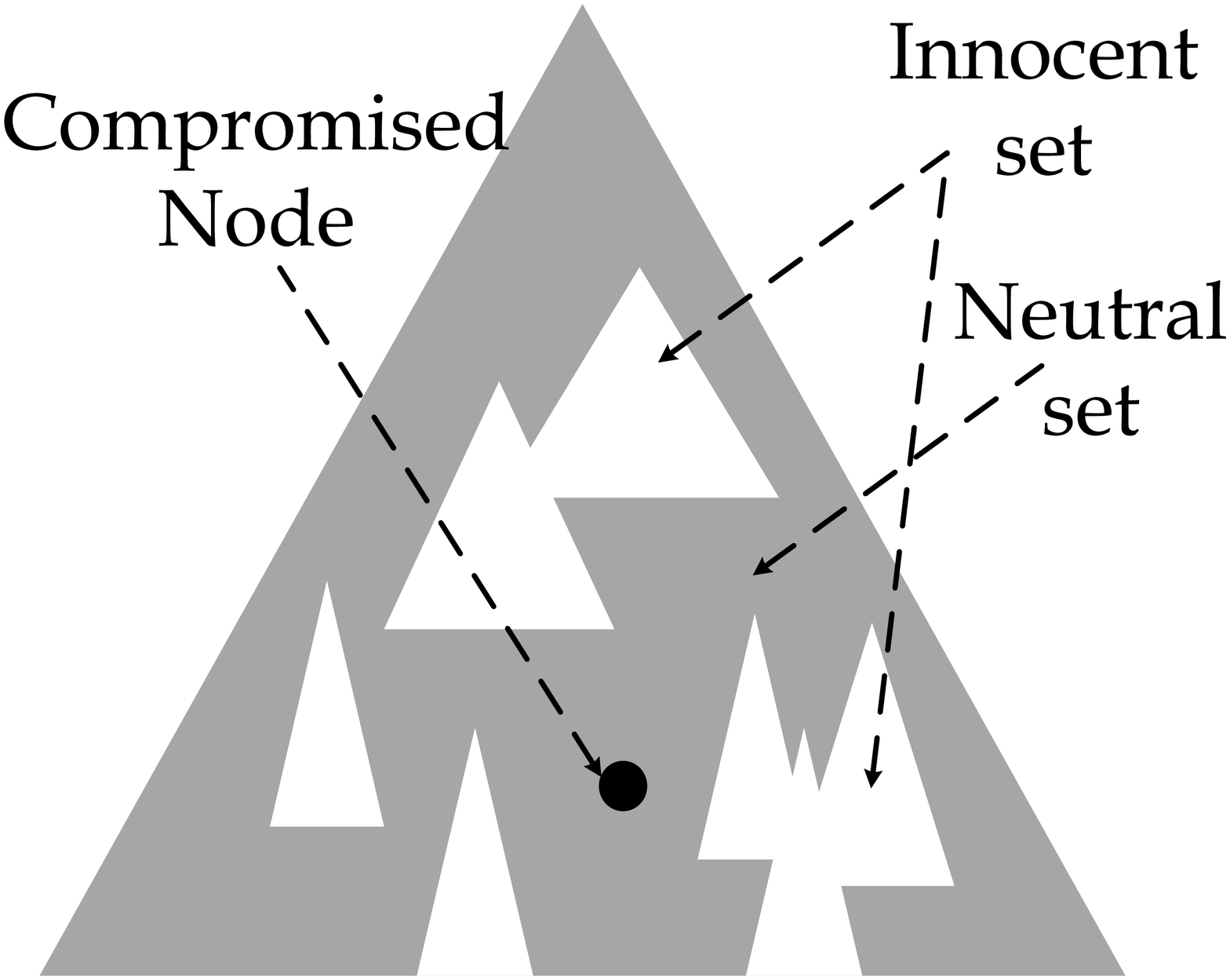}}
\caption{\scriptsize The conceptual diagrams of identifying the compromised nodes. (a) Only the nodes in red area need to be attested. (b) Only the nodes in gray area need to be attested.} \label{fig: diagram}
\end{figure}

\begin{figure}[h]
\centering
\subfloat[]{\label{fig: number of nodes to be attested N500}\includegraphics[width=0.26\textwidth]{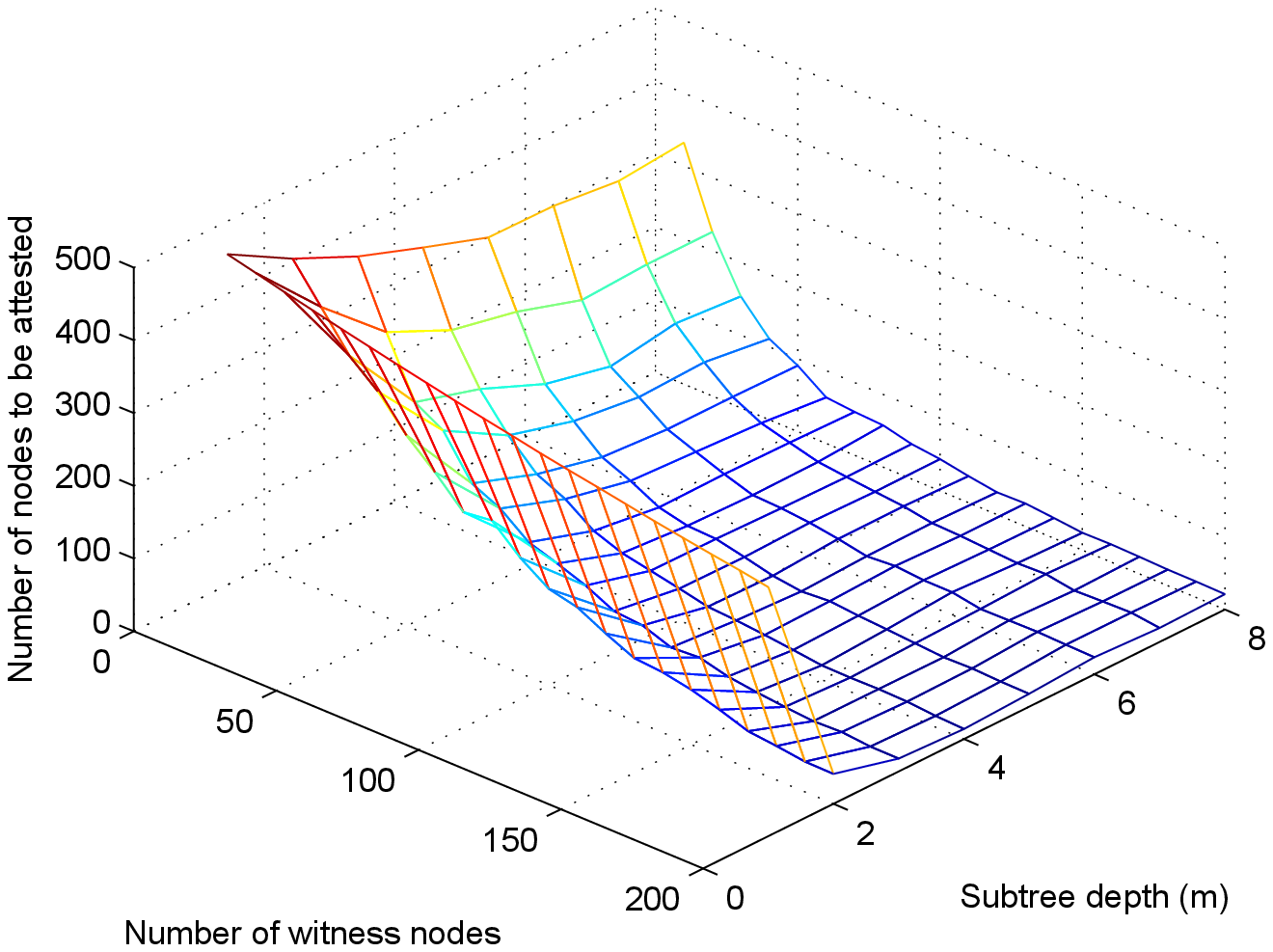}}
\subfloat[]{\label{fig: number of nodes to be attested N1000}\includegraphics[width=0.26\textwidth]{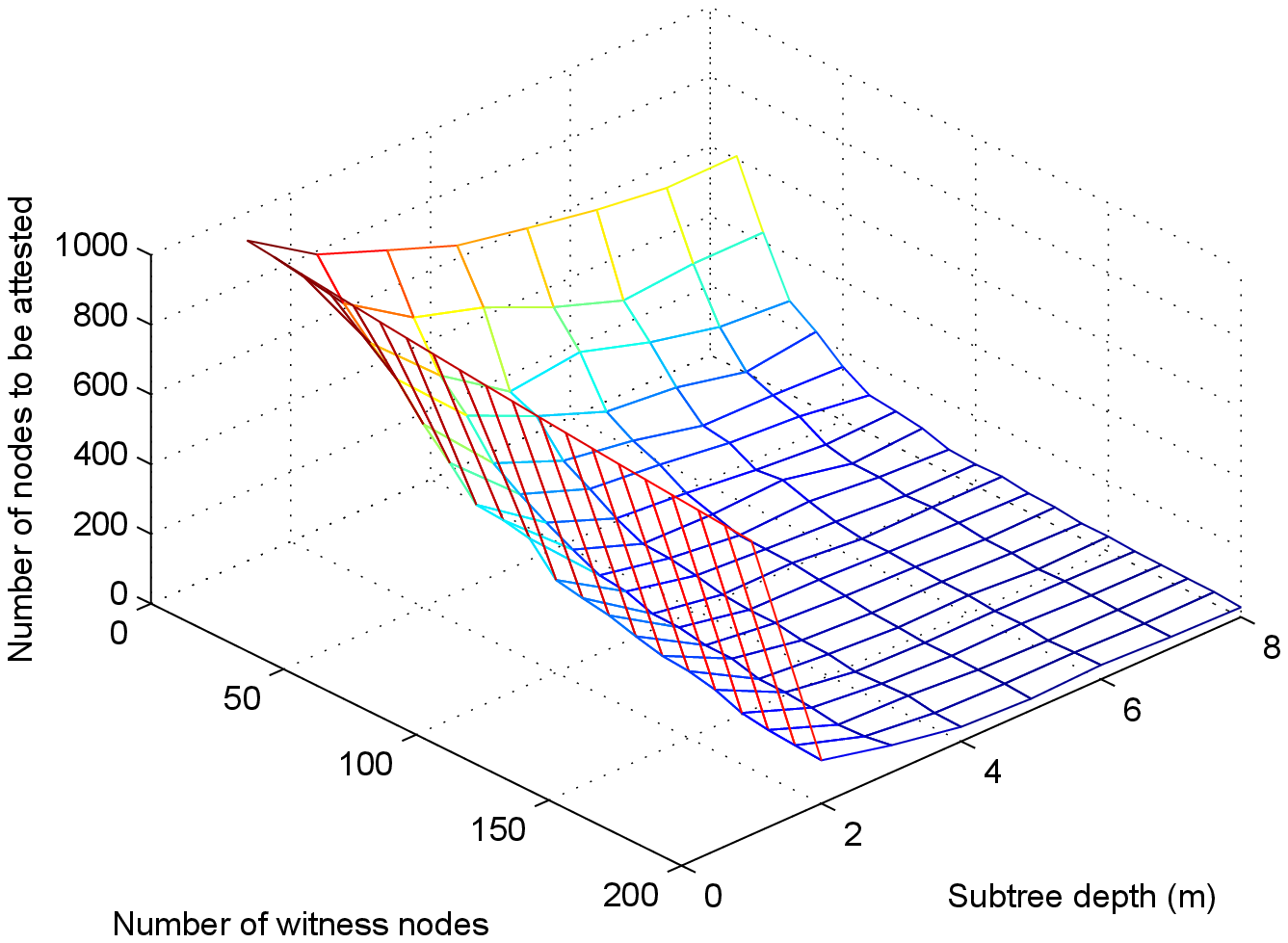}}
\caption{\scriptsize The number of nodes to be attested for (a) $N=500$ and (b) $N=1000$.} \label{fig: number of nodes to be attested}
\end{figure}

We first consider the simplest case, where no compromised nodes act as the witness nodes. We further assume that only one compromised node (\emph{i.e.}, $|\mathfrak{s}|=1$) injects false sub-proof for simplicity and our method can be adapted to the case of multiple compromised nodes injecting false sub-proofs. We have the following observation regarding each witness node $s_i$ and its generated $\vartheta_{i,t}^m$. Assume that the query-result is found to be incomplete at epoch $t$. The authority requests $\vartheta_{i,t}^m$ for each witness node $s_i$ stored in $\mathcal{M}$. To identify the compromised nodes, all the nodes at first are considered neutral. The nodes in $T_i^m$ become innocent if the verification of $\vartheta_{i,t}^m$ passes, and the nodes in $T_i^m$ become suspicious otherwise\footnote{Note that there would be the cases that a node is deemed to be both innocent and suspicious when different subtree proofs are considered. When such a case happens, that node obviously should be innocent.}. The above verification is performed as follows. According to the $\bar{P}_{i,t}^m$ in the subtree proof $\vartheta_{i,t}^m$, the authority knows which nodes in the subtree $T_i^m$ contribute data and their amount. The authority can, based on this information, calculate $\bar{H}_{i,t}^m$ by itself. Define $\mathcal{E}_{T_i^m,t}$ as the set of data sensed by the nodes in $T_i^m$, and $\mathcal{B}_{T_i^m,t}$ as the corresponding bucket IDs of $\mathcal{E}_{T_i^m,t}$. As a consequence, the verification passes if and only if the $\bar{H}_{i,t}$ calculated by the authority itself is equal to the $\bar{H}_{i,t}$ extracted from the received received $\vartheta_{i,t}^m$, and the amount of the data in $\mathcal{E}_{T_i^m,t}$ falling into the specified buckets in $\mathcal{B}_{T_i^m,t}$ matches the one indicated in $\bar{P}_{i,t}$. As a whole, each time the above checking procedures are performed according to $\vartheta_{i,t}^m$ at epoch $t$, nodes will be partitioned into three sets, innocent set $\mathfrak{I}_t^i$, neutral set, and suspicious set $\mathfrak{S}_t^i$ containing innocent nodes, neutral nodes, and suspicious nodes, respectively. We can conclude that $(\bigcap_{i\in W_t,\mathfrak{S}_t^i\neq \emptyset}\mathfrak{S}_t^i)\bigcup\{\mathcal{M}\}$ contains at least one compromised node injecting the false subproof if at least one $\mathfrak{S}_t^i$ is nonempty and $(\{s_1,\dots,s_{N-1}\}\setminus(\bigcup_{i\in W_t}\mathfrak{I}_t^i))\bigcup\{\mathcal{M}\}$ contains at least one compromised node injecting the false subproof otherwise. In more details, the authority at first only performs the attestation \cite{slpdk06,smkk05,spdk04} on the nodes in $\bigcap_{i\in W_t,\mathfrak{S}_t^i\neq \emptyset}\mathfrak{S}_t^i$ or in $\{s_1,\dots,s_{N-1}\}\setminus(\bigcup_{i\in W_t}\mathfrak{I}_t^i)$. Nonetheless, after the attestation, if the nodes being attested are ensured to be not compromised, then $\mathcal{M}$ should be the compromised node. It can be observed that the size of the set of the nodes the authority needs to perform the attestation has possibility of being drastically shrunk so that the computation and communication cost required in the attestation will be substantially reduced as well. It can also be observed that each time the attestation is performed, at least one compromised node can be recovered. The intuition behind the checking procedure can be illustrated in Fig. \ref{fig: diagram}. In addition, Fig. \ref{fig: number of nodes to be attested} depicts the number of nodes to be attested after an incomplete reply is found in different settings. Since the number of witness nodes is approximately $\xi_2(N-1)/2^{\ell_h}$, it can be observed from Fig. \ref{fig: number of nodes to be attested} that the larger the $\xi_2$ and $m$, the lower the number of nodes to be attested. Nevertheless, when $\xi_2$ and $m$ become larger, the communication cost of the modified SRQ scheme, which will be presented later, is increased as well.

There, however, would still be the cases where the compromised nodes luckily act as witness nodes so that they can be considered innocent by sending genuine subtree proof to $\mathcal{M}$. In our consideration, this case does happen, but our technique also successfully mitigates the threat of false-incrimination attacks because the effectiveness of false-incrimination attacks is now limited within the case where some compromised nodes work as witness nodes.

Now, we have to remove the two unrealistic assumptions we made before, allowing that the bogus data can be injected and the subtree proof sent from the witness node has possibility to be maliciously altered on its way to the authority. After the removal of these two unrealistic assumptions, SRQ is still resilient against collusion attacks because the use of the proofs guarantees that the misbehavior of deleting the sensed data will be detected. Unfortunately, on the one hand, the compromised nodes injecting the bogus data can deceive the authority into accepting the falsified sensor reading. On the other hand, the compromised nodes lying on the path between $\mathcal{M}$ and witness nodes can manipulate the subtree proofs so that the compromised nodes can avoid the detection and the innocent $\mathcal{M}$ can still be falsely incriminated. In our strategy, we use the \emph{redundancy property}, which have been widely used in the design of the other security protocols in WSNs such as en-route filtering \cite{yg06,yl09,yllz04,zsjn04}, to mitigate the former threat, while we, motivated by the IP traceback technique in internet security literature, develop a new traceback technique suitable for WSNs to resist against the latter attack. Here, the redundancy property means that usually WSNs are densely deployed so that an event in the sensing region can be simultaneously detected by multiple nodes. In general, the redundancy property can be obtained by achieving the so-called $t$-coverage \cite{ht05,mkps01,wxzlpg03} and therefore, an event can be simultaneously observed by at least $t$ nodes. In the following description of our remedy to these problems, due to its similarity to our modified SRQ scheme previously mentioned, we will omit some notational details, stressing on the procedures itself.

When the redundancy property is used, in essence, our SRQ does not need to be changed. Assume for now that, the authority issues a range query and receives the query-result from $\mathcal{M}$. In addition, we also assume that all the checking procedures stated in (modified) SRQ are passed. The authority now wants to know whether the received data are falsified by and sent from the compromised nodes. Recall that each node can be aware of its geographic position. After knowing which nodes contribute the sensed data to its issued query from the query-result, the authority further acquires the encrypted data of the neighbors of those nodes from $\mathcal{M}$. Note that this can be achieved because when geographic positions of all nodes are known by the authority\footnote{As long as each node knows its position, it sends in a multihop manner its position to the authority with the MAC constructed by the key uniquely shared with the authority. This kind of operations are only performed once after the sensor deployment.}, inferring the one-hop neighbors of a specific node can be easily achieved. Finally, for each node in the query-result, the authority checks the consistency of its sensor reading and the sensor readings of its one-hop neighbors. The sensed data will be rejected as long as it is inconsistent with the sensor readings of its neighbors. The consistency checking procedure here may allow for certain measurement errors or environmental factors, which should be domain-specific and user-defined. 

Now, we turn to address another problem that the subtree proofs transmitted from the witness node to the authority could be maliciously altered by the compromised (storage or sensor) nodes. To deal with this kind of threat, we need a mechanism, by which the receiver not only can know whether the received message is modified by the intermediate nodes, but also can point out the node modifying the message if it does exist. Motivated by the IP traceback techniques, we develop a \emph{recursive traceback} mechanism. To be more specifically, each node $s_i$ on the path connecting the $\mathcal{M}$ and the witness node, after receiving $D$, where $D$ denotes the subtree proof, from its descendant node, attaches $hash_{K_{i,t}}(D)$ to $D$ and then forwards $D||hash_{K_{i,t}}(D)$ to its ascendant node on the underlying aggregation tree. Note that it is assumed that the witness node receives $D||\emptyset$. Thus, with this modification, a subtree proof received by the authority should be accompanied with $\psi$ hashes, where $\psi$ is the number of nodes (including storage nodes and the witness node itself) between a specific witness node and the authority. Here, we should note that because the topology of the aggregation tree is known by the authority, when the subtree proof is sent, the IDs of intermediate nodes except for the ID of the witness node itself do not need to be attached\footnote{The path from any sensor node in the aggregation tree to $\mathcal{M}$ is unique. The authority can infer the nodes the subtree proof traverses once it is aware of the ID of the witness node.}. Hence, when the query-result is deemed incomplete, before conducting the procedures defined in the subtree sampling technique to attest nodes, the authority checks whether the received subtree proof is maliciously altered by the intermediate node. In particular, assume that the subtree proof and its $\psi$ associated hashes, $D||h_1^D||\cdots||h_\psi^D$, where $h_i^D$, $1\leq i\leq \psi$, is the hash calculated by the node that is $(i-1)$-hop away from the witness node and $h_1^D$ is computed by the witness node itself according to its asserted subtree proof, are received by the authority. After the reception of $D||h_1^D||\cdots||h_\psi^D$, the authority checks the consistency of the hash backward; \emph{i.e.}, it first checks $h_{\psi}^D$, and then $h_{\psi-1}^D$, and so on. If such a verification can be successfully proceeded all $\psi$ hashes, then the subtree proof $D$ is considered to be intact. Otherwise, once the verification fails in, say, $h_j^D$, we can conclude that the subtree proof was altered by the corresponding node since the innocent node is not assumed to behave in such a way.

Recall that the communication cost of original SRQ is $O(N+d\log N)$. In the modified SRQ \emph{without those two unrealistic assumptions}, each node $s_i$ at epoch $t$ is required to additionally send $\vartheta_{i,t}^{0},\dots,\vartheta_{i,t}^{m-1}$, leading to $O(N)$ communication cost. At epoch $t$, approximately $\xi_2(N-1)/2^{\ell_h}$ witness nodes will send their subtree proofs to $\mathcal{M}$. Nevertheless, when the subtree proofs are sent to $\mathcal{M}$, since the additional hashes will be added, its communication cost will be $O(N\cdot (\sqrt{N}+(\sqrt{N}-1)+\cdots+1))$. As a result, the communication cost becomes $O(N^2+d\log N)$.

Basically, STQ can be regarded as a special use of SRQ. Thus, its resiliency against false-incrimination attack is the same as that of SRQ. Nevertheless, due to the nature of top-$k$ query, the injection of the falsified sensor reading from the compromised nodes will imply a significant query-result deviation. Nevertheless, due to the use of the redundancy property in resisting against false data injection, STQ is resilient against the collusion attacks and false-incrimination attacks even when the compromised nodes can inject bogus data and can manipulate the proofs. Finally, because of its similarity to the SRQ scheme, STQ has the communication overhead the same as SRQ's.  

\textbf{Impact of Sensor Node Compromise on SSQ.} Recall that two aforementioned unrealistic assumptions are made. We first consider the resiliency against false-incrimination attacks. The simplest method is to introduce a parameter $\xi_3\leq \xi_1$ so that $\mathcal{M}$ reports to the authority all the verification seeds, instead of the hash of them in original SSQ. After that, for each group, if at least $\xi_3$ out of $\xi_1$ verification seeds can be successfully verified, then the quasi-skyline data in that group are considered complete. Hence, the threat of false-incrimination attacks will be mitigated because the adversary is forced to send at least $\xi_1-\xi_3+1$ false proofs, instead of single one false proof. With even one verification seed from a specific group failed to be verified, we can conclude that at least one compromised sensor node exists in that group.

Now, we consider collusion attacks. In SSQ, both $\mathcal{M}$ and sensor nodes only know the quasi-skyline data. Recall that quasi-skyline data are not necessarily the skyline data. Thus, even when there is more than one compromised sensor node in a group, the probability of successfully dropping the skyline data is actually small. More specifically, to drop the skyline data at a specified epoch, $\mathfrak{s}$ should contain at least $\xi_3$ sensor nodes responsible for sending hash values in order to successfully forge a proof of incomplete quasi-skyline data, and at the same time should be fortunate enough to select groups whose quasi-skyline data contain skyline data\footnote{Definitely, $\mathcal{M}$ can simply drop all the sensed data. Nevertheless, under this option, it is forced to forge at least $\mu(\xi_1-\xi_3+1)$ proofs ($\mu=\sqrt{N}$ in our analysis), leading to high probability of being detected.}.

\begin{figure}[h]
\centering
\subfloat[]{\label{fig: SSQimpactNY}\includegraphics[width=0.26\textwidth]{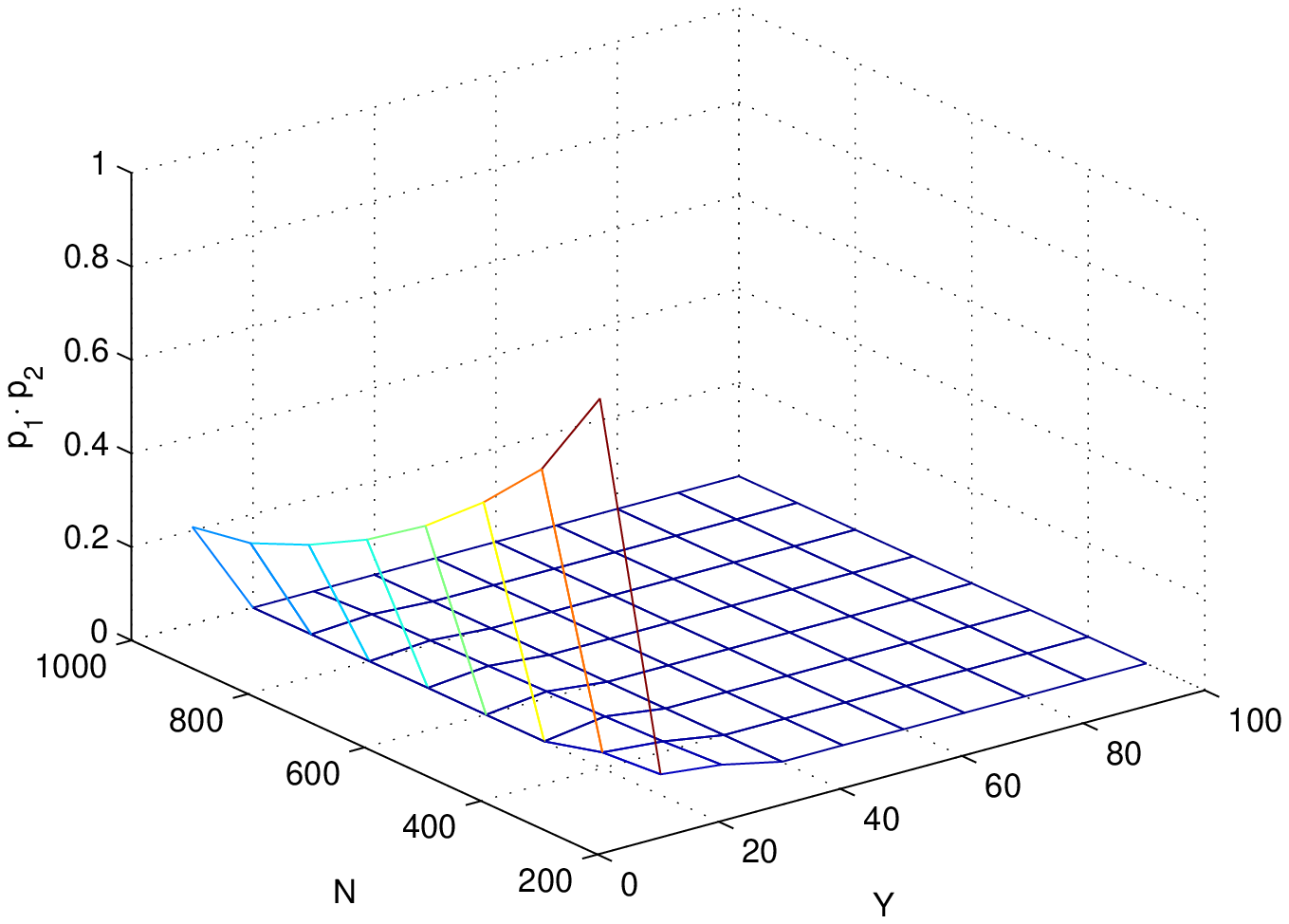}}
\subfloat[]{\label{fig: SSQimpactxi1xi3}\includegraphics[width=0.26\textwidth]{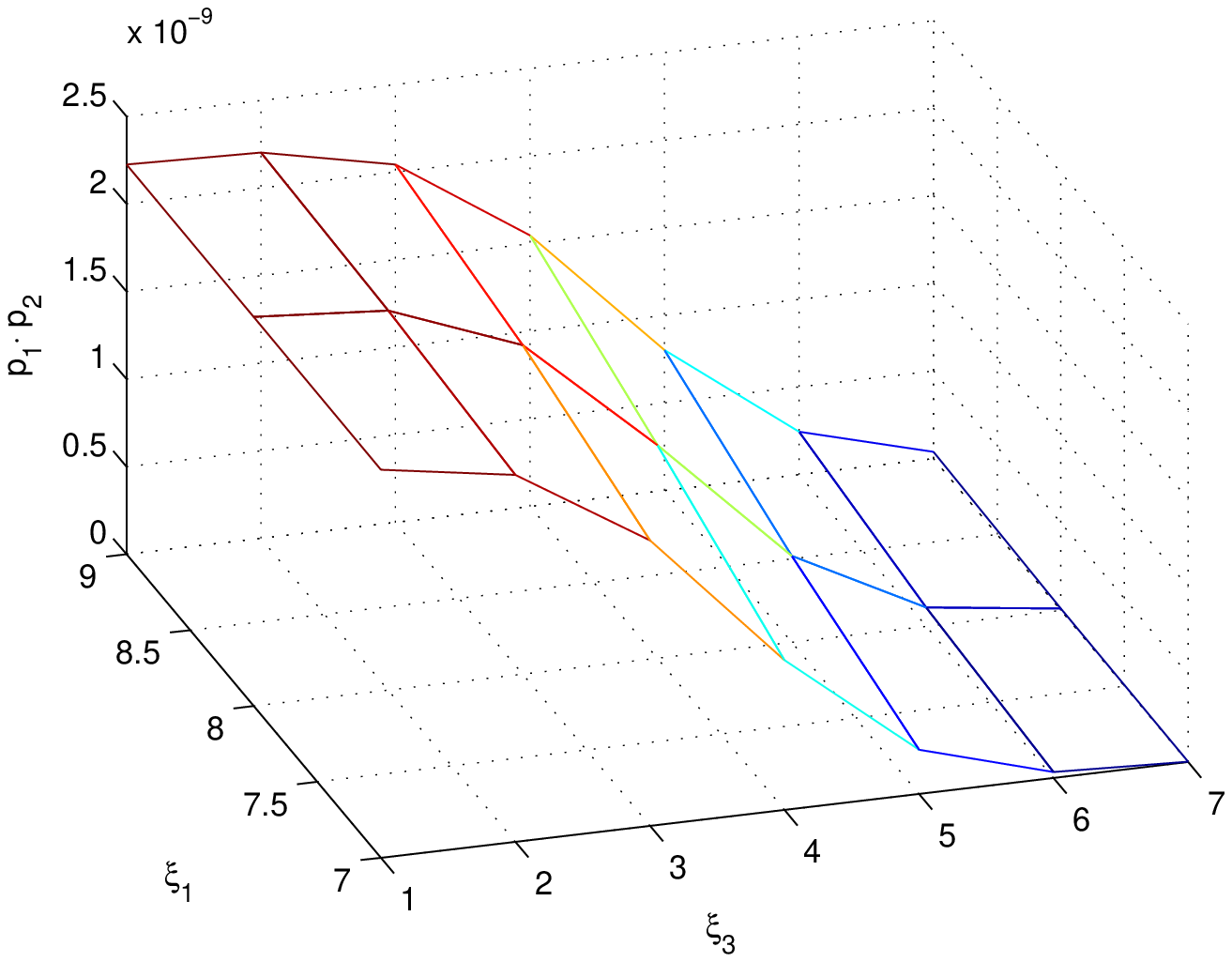}}
\caption{\scriptsize The probability of successfully dropping skyline data in the cases that (a) $\xi_1=8$ and $\xi_3=4$, and (b) $N=500$ and $Y=100$.} \label{fig: SSQimpact}
\end{figure}

Now, we consider both the collusion and false-incrimination attacks. To drop the skyline data, the only thing $\mathcal{M}$ can do is to drop the data of certain groups. Here, for simplicity, we consider the case where $\mathcal{M}$ drops the data of a fixed group $G_\eta$, $\eta\in[1,\mu]$, in which a set $\mathfrak{s}$ of $x$ sensor nodes is compromised. To prevent the detection of incomplete query-result, $\xi_3$ out of $x$ compromised sensor nodes should be the sensor nodes responsible for sending the proofs. The probability that at least $\xi_3$ out of $\xi_1$ nodes responsible for sending the proofs are contained in $\mathfrak{s}$ is $p_1=\sum_{j=\xi_3}^{\xi_1}{\xi_1\choose j}{|G_\eta|-\xi_1\choose x-j}/{|G_\eta|\choose x}$. In other words, this is equal to the probability that the compromised sensor nodes can drop the quasi-skyline data without being detected.  Quasi-skyline data, however, are not necessarily equivalent to the skyline data. The probability that the quasi-skyline data dropped by compromised nodes indeed contain skyline data can be represented as $p_2={Y-p_cY\choose |G_\eta|Y/N-p_cY}/{Y\choose |G_\eta|Y/N}$, where $p_c$ is the average ratio of the skyline data to all the sensed data. In short, this is equal to the probability that the operations performed by compromised nodes cause the loss of skyline data. As a whole, even if the adversary has $x$ compromised nodes in $G_\eta$, the probability of successfully making skyline query-result incomplete is merely $p_1\cdot p_2$. The trends of such a probability are depicted in Fig. \ref{fig: SSQimpact} under different parameter settings. As shown in Fig. \ref{fig: SSQimpactNY}, $p_1\cdot p_2$ is decreased with an increase of $N$ and $Y$. This is because when $N$ and $Y$ become larger, it is more unlikely that the quasi-skyline data dropped by the adversary contains the skyline data. Nevertheless, as shown in Fig. \ref{fig: SSQimpactxi1xi3}, $p_1\cdot p_2$ is increased with an increase of $(\xi_1-\xi_3)$. This is because when $(\xi_1-\xi_3)$ becomes larger, it is more likely that $\mathfrak{s}$ can forge a proof of an incomplete quasi-skyline data. As a whole, from the false-incrimination attack point of view, the larger the $(\xi_1-\xi_3)$, the lower the $p_1\cdot p_2$, but from the collusion attack point of view, the smaller the $(\xi_1-\xi_3)$, the lower the $p_1\cdot p_2$. This would be an optimization problem that deserves further studying. In addition, compared with original SSQ, the additional communication cost incurred from the modified SSQ comes from the transmission of verification seeds from $\mathcal{M}$ to the authority. Thus, the communication cost of the modified SSQ scheme remains $O(N^{\frac{3}{2}}+Nd)$.

Now, the two unrealistic assumptions will be relaxed so that the compromised nodes can provide falsified sensor readings and contaminate the proofs. The same as the top-$k$ query, skyline query is vulnerable to the falsified data. That is, a compromised node injecting an falsified extreme sensor reading can gain the effect of deleting all the data sensed by the other sensor nodes. Thus, the redundancy property is sill required to be applied on our SSQ scheme so as to detect the false data injection. Because the use of the redundancy property in SSQ is also similar to its use in SRQ, we omit the detailed description as well. In addition, the adversary may compromise the sensor nodes near the innocent $\mathcal{M}$ so that it can falsify all the verification seeds. Therefore, our recursive traceback mechanism is also required to be applied on SSQ to secure the verification seeds. Because of these two additional changes in SSQ, the communication cost becomes $O(N^2+Nd)$.

\section{Conclusion}\label{sec: Conclusion}
We propose schemes for securing range query, top-$k$ query, and skyline query, respectively. Two critical performance metrics, detection probability and communication cost, are analyzed. In particular, the performance of SRQ is superior to all the prior works, while STQ and SSQ act as the first proposals for securing top-$k$ query and skyline query, respectively, in tiered sensor networks. We also investigate the security impact of collusion attacks and newly identified false-information attacks, and explore the resiliency of the proposed schemes against these two attacks.

\end{document}